\documentclass{jfm}

\ifCUPmtlplainloaded \else
  \checkfont{eurm10}
  \iffontfound
    \IfFileExists{upmath.sty}
      {\typeout{^^JFound AMS Euler Roman fonts on the system,
                   using the 'upmath' package.^^J}%
       \usepackage{upmath}}
      {\typeout{^^JFound AMS Euler Roman fonts on the system, but you
                   dont seem to have the}%
       \typeout{'upmath' package installed. JFM.cls can take advantage
                 of these fonts,^^Jif you use 'upmath' package.^^J}%
       \providecommand\upi{\pi}%
      }
  \else
    \providecommand\upi{\pi}%
  \fi
\fi

\ifCUPmtlplainloaded \else
  \checkfont{msam10}
  \iffontfound
    \IfFileExists{amssymb.sty}
      {\typeout{^^JFound AMS Symbol fonts on the system, using the
                'amssymb' package.^^J}%
       \usepackage{amssymb}%
         \let\leq=\leqslant
         \let\geq=\geqslant
      }{}
  \fi
\fi

\ifCUPmtlplainloaded \else
  \IfFileExists{amsbsy.sty}
    {\typeout{^^JFound the 'amsbsy' package on the system, using it.^^J}%
     \usepackage{amsbsy}}
    {}
\fi

\newif\ifdraft
\drafttrue     

\usepackage{color,graphicx}
\usepackage{amsmath,amsfonts,amssymb}
\usepackage{natbib}
\usepackage{subfigure}

    \ifdraft    

\newcommand{\JFG}[1]{$\footnotemark\footnotetext{JG: #1}$}

\newcommand{\EB}[1]{$\footnotemark\footnotetext{JH: #1}$}

    \else   

\newcommand{\JFG}[1]{}
\newcommand{\EB}[1]{}

    \fi

\newcommand{\ie}{{i.e.}}        

\newcommand{\bx}{\ensuremath{{\bf x}}}
\newcommand{\grad}{\ensuremath{{\bf \nabla}}}
\newcommand{\bu}{\ensuremath{{\bf u}}}
\newcommand{\be}{\ensuremath{{\bf e}}}
\newcommand{\bff}{\ensuremath{{\bf f}}}
\newcommand{\ptot}{\ensuremath{p_{\text{tot}}}}
\newcommand{\butot}{\ensuremath{{\bf u}_{\text{tot}}}}
\newcommand{\utot}{\ensuremath{u_{\text{tot}}}}
\newcommand{\Reynolds}{\ensuremath{Re}}
\newcommand{\Rey}{\ensuremath{Re}}
\newcommand{\dPdx}  {\ensuremath{dP\!/\!dx}}
\newcommand{\lapl}{\ensuremath{\nabla^2}}
\newcommand{\hutotO}{\ensuremath{\hat{u}_{\text{tot},0}}}
\newcommand{\hbutotO}{\ensuremath{\hat{{\bf u}}_{\text{tot},0}}}

\newcommand{\hbu}{\ensuremath{\hat{{\bf u}}}}
\newcommand{\hu}{\ensuremath{\hat{u}}}
\newcommand{\hv}{\ensuremath{\hat{v}}}
\newcommand{\hw}{\ensuremath{\hat{w}}}
\newcommand{\hp}{\ensuremath{\hat{p}}}

\newcommand{\tbu}{\ensuremath{\tilde{{\bf u}}}}
\newcommand{\tu}{\ensuremath{\tilde{u}}}
\newcommand{\tv}{\ensuremath{\tilde{v}}}
\newcommand{\tw}{\ensuremath{\tilde{w}}}
\newcommand{\tp}{\ensuremath{\tilde{p}}}
\newcommand{\teta}{\ensuremath{\tilde{\eta}}}

\newcommand{\refeq}  [1] {(\ref{#1})}

\newcommand{\eqn}[1]{eqn.\ {\ref{#1}}}
\newcommand{\Eqn}[1]{Eqn.\ {\ref{#1}}}

\newcommand{\reffig} [1] {figure~\ref{#1}}
\newcommand{\reffigs} [2] {figures~\ref{#1} and~\ref{#2}}
\newcommand{\refFig} [1] {Figure~\ref{#1}}

\newcommand{\reftab} [1] {table~\ref{#1}}
\newcommand{\refTab} [1] {Table~\ref{#1}}

\newcommand{\refsec}[1] {\S\,\ref{#1}}

\newcommand{\tx}{\ensuremath{\tau_x}}
\newcommand{\tz}{\ensuremath{\tau_z}}
\newcommand{\txz}{\ensuremath{\tau_{xz}}}

\newcommand{\sy}{\ensuremath{\sigma_{y}}}
\newcommand{\sz}{\ensuremath{\sigma_{z}}}
\newcommand{\sxy}{\ensuremath{\sigma_{xy}}}

\newcommand{\sxyz}{\ensuremath{\sigma_{xyz}}}

\newcommand{\cc}{\text{c.c.}}

\usepackage{graphicx}
\usepackage{natbib}
\usepackage{dcolumn}
\usepackage{multirow}
\usepackage{hyperref}

\newif\ifcolorfigs
\colorfigstrue     

\ifcolorfigs            
  \graphicspath{{./figs-clr-small/}{./figs/fig-clr/}}
  \newcommand{\colorcomm}[2]{#1}
\else
  \graphicspath{{./figs-bw-small/}{./figs/fig-bw/}}
  \newcommand{\colorcomm}[2]{#2}
\fi

\title[Spatially localized solutions of shear flows]
{Spatially localized solutions of shear flows}

\author[
J.F.\ Gibson
    and
E.\ Brand
]
{
J.\ns F.\ns G\ls I\ls B\ls S\ls O\ls N
\and
E.\ns  B\ls R\ls A\ls N\ls D
}

\affiliation{
Dept.\ of Mathematics and Statistics, University of New Hampshire,
Durham, NH 03824, USA
}

\date{\today}
\begin{document}
\maketitle

\begin{abstract}
We present several new spatially localized equilibrium and traveling-wave solutions of plane 
Couette and channel flows. The solutions exhibit strikingly concentrated regions of vorticity 
that are flanked on either side by high-speed streaks. For several traveling-wave solutions 
of channel flow, the concentrated vortex structures are confined to the near-wall region and 
form particularly isolated and elemental versions of coherent structures in the near-wall 
region of shear flows. 
The solutions are constructed by a variety of methods: application of windowing functions 
to previously known spatially periodic solutions, continuation from plane Couette to channel 
flow conditions, and from initial guesses obtained from turbulent simulation data. 
We show how the symmetries of localized solutions derive from the symmetries of their 
periodic counterparts, analyze the exponential decay of their tails, examine the scale 
separation and scaling of their streamwise Fourier modes, and show that they develop 
critical layers for large Reynolds numbers.

\end{abstract}

\vspace*{-0.6cm}

\section{Introduction}

Over the last twenty years the computation of invariant solutions of the Navier-Stokes
equations, or ``exact coherent structures,'' has opened a new approach to 
understanding the dynamics of moderate-Reynolds unsteady flows, an approach which 
promises to provide a long-hoped-for bridge between dynamical systems theory and 
turbulence. Unlike previous derivations of low-order dynamical models of unsteady flows 
(\cite{LorenzJAS63, AubryJFM88, Holmes96}), the invariant-solutions approach forgoes 
low-d projections and simplified models and instead takes a fully-resolved direct 
numerical simulation as a quantitatively accurate finite-dimensional approximation 
of the Navier-Stokes equations. The fully-resolved simulation is then treated as 
very high-dimensional dynamical system. 
The first step in analysis of a dynamical system is the computation of its invariant 
solutions: its equilibria, homo- and heteroclinic orbits, and periodic orbits. In the 
dynamical systems view of turbulence, the equilibrium solutions correspond to steady 
states of the fluid flow, periodic orbits correspond to states of the fluid velocity 
field repeat themselves exactly after a finite time, and homo- and heteroclinic orbits 
correspond to dynamic transitions between equilibria or periodic orbits. For flows
with homogeneous spatial directions, such as pipes and channels, continuous symmetries 
in the equations of motion allow relative invariant solutions, e.g. traveling waves 
(relative equilibria) and relative periodic orbits. 

The simplest invariant solutions of fluids are the classical, closed-form steady 
states of the Navier-Stokes equations, for example, the parabolic laminar flow 
profile of pressure-driven channel and pipe flow, or the linear laminar solution
of plane Couette flow. Solutions such as these have special cancellations which 
make it possible to represent the exact solution of the nonlinear system with a finite 
set of simple functions. For example, for the laminar solution of channel flow, the nonlinear 
term vanishes and the solution can be represented exactly as a 2nd order polynomial 
in the wall-normal variable. However, if we consider the problem from the perspective 
of faithful, very high-dimensional finite discretizations, invariant solutions 
are the solutions of a nonlinear algebraic or differential equations in 
$O(10^4)$ to $O(10^6)$ free variables. Only a few very special solutions 
(the classical ones) will involve few enough modes to be expressible in closed form,
and most will involve nonlinear coupling between large numbers of nonzero variables.
Compared to the classical closed-form solutions, these computed invariant solutions 
are typically unstable, fully three-dimensional, fully nonlinear, distant from the 
smooth laminar flow solutions, and involve most if not all the available modes of 
the numerical representation. Determination and specification of such solutions is 
necessarily numerical.

The practical feasibility of finding such high-dimensional nonlinear solutions 
of Navier-Stokes was first demonstrated by \cite{NagataJFM90}, who computed 
an unstable 3-dimensional nonlinear equilibrium solution of plane Couette flow at a Reynolds 
number above the onset of turbulence, using a 589-dimensional discretization. 
The same equilibrium solution was found independently and analyzed in greater 
precision and detail by \cite{CleverJFM92} and \cite{WaleffePRL98,WaleffePF03}. 
A large number of equilibria and traveling waves of plane Couette and pipe flow
have since been found (\cite{NagataPRE97,SchmiegelPHD99,FaisstPRL03,GibsonJFM09}), 
a few of channel flow (\cite{WaleffeJFM01,ItanoJPSJ01}), and in other flows such 
square duct flow (\cite{OkinoJFM10}, \cite{WedinPRE09}). Periodic orbits have been 
calculated for plane Couette flow (\cite{KawaharaJFM01,ViswanathJFM07,CvitanovicPS10}),
pipe flow (\cite{DuguetPF08}), and 2D Kolmogorov turbulence (\cite{ChandlerJFM13}), 
and hetero- and homoclinic connections
for plane Couette flow (\cite{GibsonJFM08,HalcrowJFM09,VanVeenPRL11}). Improved 
numerical methods and more powerful computers now allow the computation of 
solutions with as many as $10^6$ free variables. High-resolution calculations 
have shown that discretization errors converge toward zero as resolution is 
increased, demonstrating that the numerical solutions are precise approximations
of true solutions of the Navier-Stokes equations, rather 
than artifacts of discretization. High-resolution calculations have also allowed 
accurate computation of solutions with fine spatial structure, such as periodic 
orbits that exhibit turbulent ``bursting'' phases (\cite{ViswanathJFM07,CvitanovicPS10}). 

Just as in low-dimensional dynamical systems theory, the importance of these 
invariant solutions stems from the organization they impose on the state space 
dynamics. In particular, dynamics in the neighborhood of (relative) equilibria 
and periodic orbits is governed to leading order by the linearization about these solutions,
and the eigenvalues of the linearized dynamics reveal the local character of the 
state space flow and the dimensionality of each solution's unstable manifold. 
For shear flows at moderate Reynolds numbers and in closed or small periodic 
domains, most known invariant solutions have a positive but remarkably 
small number of unstable eigenvalues, and correspondingly low-dimensional unstable 
manifolds. For example, the equilibrium solution of plane Couette flow developed by 
Nagata, Busse, Clever, and Waleffe (hereafter termed the NBCW equilibrium) has a 
{\em single} unstable eigenvalue (\cite{WangPRL07}, and the periodic orbit solutions 
of \cite{ViswanathJFM07} have between 1 and 11 unstable eigenvalues. This low 
dimensionality of instability is a crucially important result. It suggests, as long suspected, 
that moderate-Reynolds flows are inherently low-dimensional --at least for small
confined domains. It further suggests that the temporal dynamics of such flows
results from a relatively low-dimensional, chaotic but deterministic walk between 
the flow's unstable invariant solutions, along the low-dimensional 
network of their unstable manifolds (\cite{GibsonJFM08}). Moreover, the coherent 
structures often observed in such flows can be understood as resulting from close 
passes to these unstable invariant solutions, at which the Navier-Stokes equations 
balance exactly --an idea well-expressed by Waleffe's term ``exact coherent structures'' 
(\cite{WaleffeJFM01}). Indeed, a key feature of the NBCW invariant solution is that 
it captures structure commonly observed in shear flows in the form of wavy rolls 
that support alternating streaks of high and low streamwise velocity (see 
\refsec{s:periodic_pcf_solns} for further discussion). We refer the reader to the 
\cite{KawaharaARFM12} review article for an excellent overview of research 
in this area.

On the other hand, a significant weakness of the invariant-solutions 
approach to date is the assumption of idealized computational domains, 
for example, small periodic ``minimal flow units'' (\cite{HamiltonJFM95}), 
just large enough to contain a single coherent structure. For example, 
most of the work analyzing dynamics via invariant solutions has been done
in plane Couette flow in doubly-periodic boxes with stream- and spanwise 
aspect ratios around 2 or 3. Small periodic or confined domains are 
an understandable simplifying assumption for the demanding
computational problem of computing invariant solutions. However, it is also valid 
to criticize such assumptions as unphysical and limiting the research to 
explorations of {\em temporal} complexity of periodic spatial structures, 
as opposed to the full {\em spatio-temporal} complexity of turbulence in 
extended domains. The assumption of periodicity also complicates coordination 
of theory and experiment. Close passes to unstable traveling waves with axial 
periodicity have been have been detected in experimental pipe flows 
(\cite{HofScience04,LozarPRL12}), but the effort to match experiment
and theory would be greatly aided if the assumption of periodicity in 
the computations could be relaxed. Lastly, the most prominent example of 
coherent structures in fluids, and the initial motivating problem for
dynamical-systems approaches, are the lambda and hairpin vortices that form 
in the high-shear region near the walls of channel and boundary layer flows.
These appear in coupled systems of coherent structures that are individually 
localized in the wall-normal direction as well as the two 
homogeneous directions (\cite{AdrianPF07}).

The first step in addressing the weakness of small, idealized domains is to 
compute localized invariant solutions on extended domains. 
Several papers have made promising steps in this direction.
\cite{SchneiderJFM09} computed the first spatially localized equilibrium 
and traveling waves of Navier-Stokes, by an ``edge-tracking'' algorithm 
for plane Couette flow in a streamwise-periodic but spanwise-extended domain,
and showed that these solutions were spanwise-localized forms of the spatially
periodic NBCW solution, which exhibit exponential decay towards laminar flow 
in the spanwise coordinate. Another computation 
in the same paper produced an unsteady, time-varying state with exponential
localization in both span- and streamwise directions. However, this state 
appears to wander chaotically and so is not an invariant solution. A 
similar time-varying doubly-localized state was found by \cite{DuguetPF09}. 
\cite{SchneiderPRL10} demonstrated a number of interesting connections 
between the localized solutions of \cite{SchneiderJFM09} and localized
solutions of the Swift-Hohenberg equation. Both systems exhibit 
{\em homoclinic snaking}, a process by which localized solutions 
grow additional structure at their fronts via a sequence  
of saddle-node bifurcations in a continuation parameter (see 
\refsec{s:scaling} for further discussion). This is an intriguing 
connection, as Swift-Hohenberg is a key model equation in the theory 
of pattern formation (\cite{Hoyle06}), for which localization is comparatively
well-understood (\cite{BurkeChaos07}). 

The broad purposes of the present paper are (1) to further extend the 
invariant-solutions approach to turbulence to spatially extended flows,
towards the long-term goals of addressing spatio-temporal complexity  
in turbulence and aiding the effort to verify invariant solutions
in experiment, and (2) to begin an effort to capture spatially isolated 
coherent structures that occur near the walls of shear flows.
The specific results and organization of the paper are as follows. 
In \refsec{s:planecouette} we show that the spanwise localized 
solutions of \cite{SchneiderPRL10,SchneiderJFM09} are not anomalous, but that 
localized versions of other spatially periodic solutions exist and can be 
constructed easily by a windowing and refinement method that, unlike 
edge-tracking, puts no restrictions on the number of the solution's unstable 
eigenmodes. We show how the symmetries of localized solutions result from the 
symmetries and phase of the underlying periodic solution. 
In \refsec{s:channelsolns}, to further develop the invariant-solutions approach 
in experimentally accessible flow conditions,  we construct localized traveling 
wave solutions of channel flow, by windowing and refining periodic solutions 
obtained by continuation from plane Couette conditions and by searching among 
turbulent simulation data. In doing so we find particularly intriguing 
traveling-wave solutions of channel flow whose vorticity is concentrated 
in the near-wall region, in spanwise and wall-normal localized structures 
that resemble lambda vortices in streamwise-developing channel flows. 
In \refsec{s:discussion} we analyze the tails of the localized solutions and 
show that they decay exponentially to laminar flow at the rate determined solely 
by the streamwise wavenumber of the solution, with far-field structure that is 
independent of the details of the core region. We examine scale separation and 
scaling in the streamwise Fourier harmonics and development of critical layers 
at large Reynolds numbers.

\section{Equilibrium solutions of plane Couette flow}
\label{s:planecouette}

\subsection{Equations of motion}
\label{s:eqns_PCF}

Plane Couette flow consists of an incompressible fluid confined between two parallel 
rigid plates moving in-plane at a constant relative velocity. The $\bx = (x,y,z)$ 
coordinates are aligned with the streamwise, wall-normal, and spanwise directions, 
where streamwise is defined as the direction of relative wall motion. We assume a 
computational flow domain $\Omega = [-L_x/2, \: L_x/2] \times [-h, h] \times [-L_z/2, \: L_z/2]$ 
with periodic boundary conditions in $x$ and $z$ and no-slip conditions at the walls 
$y= \pm h$. We restrict our attention to streamwise-periodic velocity fields and $L_x$
chosen to match the streamwise wavelength. In the spanwise direction, we choose $L_z$ 
either to match the spanwise wavelength of a spanwise periodic field, or to a large 
value that approximates a spanwise-infinite domain. We decompose the total velocity 
and pressure fields into sums of a laminar 
base flow and a deviation from laminar: $\butot(\bx,t) = \bu(\bx,t) + U(y) \, \be_x$ 
and $\ptot = p(\bx,t) + x \; \dPdx$, where $\dPdx$ is a fixed constant specifying an 
imposed mean pressure gradient. For plane Couette flow we will consider only the
case $\dPdx = 0$, for which the laminar solution is $U(y) =  \bar{U} y/h$, where
$\bar{U}$ is half the relative wall speed. After nondimensionalization by $\bar{U}$,
$h$, and the kinematic viscosity $\nu$, the Navier-Stokes equations for plane Couette 
flow can be written
\begin{equation} 
\frac{\partial \bu}{\partial t} + U \frac{\partial \bu}{\partial x} + v \: U'\: \be_x  
 + \bu \cdot \grad \bu = -\nabla p  + \frac {1} {\Rey} \nabla^2 \bu, \quad \nabla \cdot \bu  = 0 
\label{eq:NSE_PCF}
\end{equation}
where $\Rey = \bar{U} h/\nu$ and the velocity components are $\bu(\bx,t) = [u,v,w](x,y,z,t)$.
In this decomposition the plane Couette laminar solution is $U(y) = y$, $\dPdx = 0$, 
$\bu = 0$, and $p = 0$. From here on we refer to $\bu$ as velocity and $\butot$ as 
total velocity, and we note that $\bu$ has Dirichlet boundary conditions at the walls. 

The Navier-Stokes equations \refeq{eq:NSE_PCF} with plane Couette conditions 
and $y$-Dirichlet, $x,z$-periodic boundary conditions are invariant under any 
combination of rotation by $\upi$ about the $z$ axis, reflection about the $z=0$ plane, 
and finite translations in the $x$ and $z$ directions. The generators of the plane Couette 
symmetry group are thus 
\begin{align}
\sigma_{xy} &: [u,v,w](x,y,z) \rightarrow [-u,-v,w](-x,-y,z) \nonumber\\
\sigma_z    &: [u,v,w](x,y,z) \rightarrow [u, v,-w](x,y,-z)  \label{PCF_gens}\\
\tau(\Delta x, \Delta z) &: [u, v, w](x,y,z) \rightarrow [u, v, w](x+\Delta x, y, z+\Delta z). \nonumber
\end{align}
We use the standard group-theory notation $\langle \ldots \rangle$ to indicate 
the group generated by a set of group elements; thus the symmetry group of plane 
Couette flow is $\langle \sxy, \sz, \tau(\Delta x, \Delta z) \rangle$. For each subgroup of this group, 
there is a subspace of velocity fields that is invariant under the equations of motion. 
That is, if a velocity field $\bu(\bx,0)$ satisfies $\bu = \sigma \bu$ for each symmetry 
$\sigma$ in a given subgroup, $\bu(\bx,t)$ will satisfy the same symmetries for all time. 
Invariant solutions of the equations of motion naturally lie in these subspaces. For 
example, with appropriate choice of the origin, equilibrium solutions of plane Couette 
typically satisfy $[u,v,w](x,y,z) = [-u, -v,-w](-x,-y,-z)$ or related symmetries involving 
$\sxy$ and $\sz$, since these symmetries requires the velocity field to vanish at the origin, 
and so prevent drifting in $x$ and $z$.

\subsection{Spatially periodic solutions: EQ1/NBCW, EQ7/HSV, and EQ8}
\label{s:periodic_pcf_solns}

\begin{figure}
\begin{center}
\begin{tabular}{ccc}
\hspace{-1.8mm} {\footnotesize (a)} \hspace{-1.8mm} \includegraphics[width=0.30\textwidth]{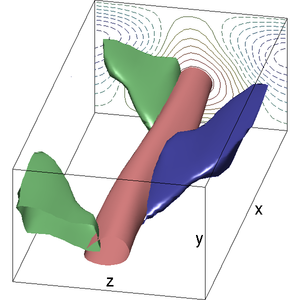} 
\hspace{-1.8mm} {\footnotesize (b)} \hspace{-1.8mm} \includegraphics[width=0.30\textwidth]{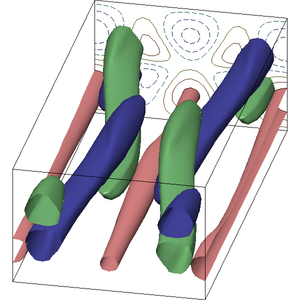} 
\hspace{-1.8mm} {\footnotesize (c)} \hspace{-1.8mm} \includegraphics[width=0.30\textwidth]{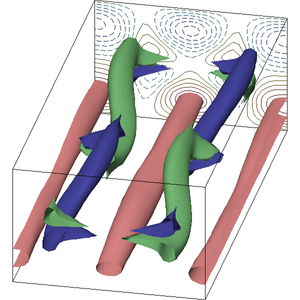}
\end{tabular}
\end{center}
\caption{{\bf Three spatially periodic equilibria of plane Couette flow.}
(a) NBCW lower branch, (b) EQ7 lower branch, and (c) EQ8 (upper branch of EQ7), all at 
$\Rey=400$ and $\alpha,\gamma = 1,2$, where $\alpha$ and $\gamma$ are the streamwise and 
spanwise wavenumbers. The visualizations show 3D isosurfaces  
of signed swirling strength at $s = \pm 0.09$ in \colorcomm{green/blue}{light/medium grey}
(see text). Isosurfaces of streamwise velocity indicating high-speed streaks are shown in 
\colorcomm{red}{dark grey} at (a) $u=0.3$, (b) $u=0.15$, and (c) $u=0.2$.
The back plane shows contours of streamwise velocity at levels (a,c) $\pm 0.03 \cdot[1\!:2\!:15]$ 
and (b) $\pm 0.03 \cdot [1\!:2\!:5]$ in Matlab notation with dashed/solid 
lines indicating negative/positive values. Note that by symmetry both solutions have streaks 
of equal magnitude and opposite streamwise velocity; these are not shown in the isosurfaces 
to reduce clutter, but they are indicated by the negative/positive symmetry of the backplane 
contours of streamwise velocity. The origin is at the center of the box.
} 
\label{f:EQ1_EQ7}
\end{figure}

\refFig{f:EQ1_EQ7} shows a visualization of three spatially periodic equilibrium 
solutions of plane Couette flow: the well-known ``lower branch'' solution of \cite{NagataJFM90}, 
\cite{CleverJFM92}, and \cite{WaleffePRL98} (NBCW, called EQ1 in \cite{GibsonJFM09})),
and the ``Hairpin Vortex Solution (HVS)'' of \cite{ItanoPRL09} and discovered 
independently as EQ7 in \cite{GibsonJFM09}. Henceforth we refer to these as  
NBCW and EQ7. EQ8 is the upper branch of the the HSV/EQ7 solution. The NBCW solution is well-known 
not only as the first known exact nonlinear solution to the Navier-Stokes equations, 
but also for a number of remarkable characteristics, which we outline briefly here. 
The NBCW solution captures precisely, in the context of plane Couette flow, the 
roll-streak structure that seems ubiquitous in shear flows ranging from Taylor-Couette 
to the turbulent boundary layer, and consequently forms an example of an exact 
instantaneous balance between the three cycles of Waleffe's self-sustaining 
process for shear flows (\cite{WaleffePF97,WaleffePRL98}).
The NBCW solution has also served as a starting point for 
recent efforts to formulate a dynamical-systems theory of turbulence. At Reynolds
numbers above the onset of turbulence, the NBCW solution lies between the laminar 
solution and the chaotic turbulent region of state space. It has a single unstable 
eigenvalue (\cite{WangPRL07}, so that its stable manifold forms a boundary between states that decay 
to laminar flow and states that grow to turbulence (\cite{SchneiderPRE08}). The comparatively low viscous 
shear rate of the NBCW solution suggests that it might be feasible to implement a 
control strategy to stabilize this single unstable direction and obtain savings in 
the wall driving force compared to the turbulent flow (\cite{WangPRL07}). The NBCW solution also has been 
shown to have well-defined asymptotic structure in the limit of large Reynolds 
number, which can be exploited to form a reduced system that accurately captures
the structure of the solution over a wide range of Reynolds numbers (\cite{WangPRL07,HallJFM10}). The asymptotic 
structure and reduced system are particularly relevant for this work, since it seems 
likely that any analytic understanding of localization in Navier-Stokes will be more 
easily developed in the context of a reduced system. 

The EQ7 solution has been conjectured to be related to hairpin vortices 
frequently observed in the turbulent boundary layer (\cite{ItanoPRL09}). The
hairpin shapes in the visualizations of \cite{ItanoPRL09} are certainly suggestive;
however, some caution should be in order, as these visualizations of vortex 
lines cannot be compared directly to visualizations that highlight the magnitude
of vorticity, such as Q criterion, lambda criterion, and swirling strength. 
\refFig{f:EQ1_EQ7} shows NBCW, EQ7, and EQ8 visualized with signed swirling strength 
isosurfaces to show the roll structure and streamwise velocity isosurfaces
to show high-speed streaks. The swirling strength at $\bx$ is defined as 
the magnitude of the complex part of the eigenvalue of the velocity gradient 
tensor $\grad \bu(\bx)$ (\cite{ZhouJFM99}). We chose swirling strength over other 
measures of fluid circulation such as Q criterion because it most clearly 
identified in 3D isosurfaces the regions of highly concentrated circulation 
that are apparent in 2D quiver plots such \reffig{f:couette_crosssections}. 
Since the invariant solutions in this paper have elongated regions of 
concentrated circulation nearly aligned with the $x$ axis, we attached a $\pm$ 
sign to the swirling strength that indicates clockwise/counterclockwise 
circulation with respect to the positive $x$ axis.

\subsection{Construction of localized initial guesses by windowing}
\label{s:windowing}

The localized equilibria and traveling waves of plane Couette flow described
in \cite{SchneiderJFM09} and \cite{SchneiderPRL10} (hereafter SGB10) are 
spanwise-localized versions of the spatially periodic NBCW solution. 
These localized solutions are comprised of a core region that closely resembles the 
periodic NBCW solution, weak tails that decay exponentially towards laminar flow, 
and a transitional region between the core and tails. This form suggests that new 
localized solutions might be found by imposing a similar core-transition-tail 
structure on other known spatially periodic solutions, and then refining these 
initial guesses with a trust-region Newton-Krylov solver. The rough form of this 
desired structure can be imposed on initial guesses by multiplying a known
spatially periodic solution, expressed as a perturbation on laminar flow, by 
an even positive windowing function $W(z)$ that is nearly unity over a core region 
$|z| < a$, decreases smoothly and monotonically to nearly zero over a transition 
region $a \leq |z| < a+b$, and vanishes as $|z| \rightarrow \infty$, followed by
projecting the resulting field $W(z) \bu(\bx)$ onto the divergence-free subspace. 
We have found that with a robust trust-region Newton-Krylov solver, the precise 
details of the windowing function and the projection are unimportant, and 
that the only important details are smoothness and the widths of the core and 
transition regions. One choice that suffices is the windowing function 
\begin{equation}
W(z) = \frac{1}{4} \; \left(1 + \tanh \left(\frac{6(a-z)}{b} + 3\right)\right) 
                      \left(1 + \tanh \left(\frac{6(a+z)}{b} + 3\right)\right).
\label{eq:window}
\end{equation}
This $W(z)$ behaves as desired: it is even, smooth, monotonic in $|z|$, 
satisfies $0.995 < W(z) < 1$ for $|z| < a$ and $0 < W(z) < 0.005$ for 
$|z| > a + b$, and it approaches zero exponentially as $|z| \rightarrow \infty$.
$W(z)$ is specified in this particular form because we found that the most 
important factor in producing a good initial guess was the size and location of the 
transition region, which are specified by the parameters $a$ and $b$. 
A sufficient projection is to apply $W(z)$ to the streamwise and spanwise 
components of velocity and reconstruct the wall-normal from the divergence-free 
condition. That is, let $\bu = [u,v,w]$ be a $z$-periodic solution expressed 
as a perturbation over laminar flow. An initial guess for a $z$-localized 
solution $\bu_g = [u_g, v_g, w_g]$ can be constructed by setting $u_g = W u$, 
$w_g = W w$, and reconstructing $v_g$ from $\grad \cdot \bu_g =0$ and boundary 
conditions. 

It is worth emphasizing that this localization procedure is rather crude. By 
construction, the initial guess should nearly satisfy the Navier-Stokes 
equations in the core region and the tails --nearly but not exactly because 
the guess merely approaches the laminar solution for large $z$, and because 
the nonlocal effects of the divergence-free projection and pressure will 
corrupt the balance of terms that one would otherwise expect in the core region 
where $W(z)$ is very nearly unity.
In the transition region, however, there is no reason to expect that the velocity 
field that smoothly interpolates between tails and core will be close to 
satisfying Navier-Stokes. The quality of these initial guesses, thus, depends 
entirely on the robustness of the solver used to refine the initial guess 
into a solution. In particular, within the transition region
the initial guess is too far from satisfying Navier-Stokes to be refined to an 
exact solution with a straight Newton method. A so-called globally convergent 
search method is instead required. We use a ``hookstep'' trust-region modification
of the Newton search method coupled with a Krylov-subspace method (GMRES) for
solution of the Newton-step equations, following \cite{ViswanathJFM07,ViswanathPTRSA09}.

\subsection{Localization and symmetry}
\label{s:localsymmetry}

The symmetries of a desired solution are important both in determining solution type
(e.g. equilibrium versus traveling wave) and for reducing the search space, which 
improves the speed and robustness of the search. The appropriate symmetries for spanwise-localized
solutions are determined as follows. We begin with a spanwise periodic
solution with a known set of symmetries. Multiplying that solution by a 
nonperiodic windowing function $W(z)$ breaks any of these symmetries that involve 
$z$ periodicity. Symmetries that do not involve $z$ periodicity are preserved 
through the localization procedure and form the symmetry group of the localized 
guess. Note further the symmetry group $G$ of a periodic solution $\bu$ 
transforms by conjugation to $\tau G \tau^{-1}$ when the solution is
phase-shifted to $\tau \bu$, and that the localization procedure will break
and preserve the different symmetry groups of $\bu$ and $\tau \bu$ differently. Thus it is possible to 
construct localized guesses with different symmetry groups by applying the windowing 
function to the same periodic solution in different spatial phases. 

To illustrate, we show how the symmetries 
of the equilibrium, traveling wave, and rung solutions of PCF in \cite{SchneiderPRL10} 
arise from localizing the NBCW solution in different spatial phases. For compactness in 
what follows let $\tx = \tau(\ell_x/2,0), \tz = \tau(0,\ell_z/2),$ and $\txz = \tx \tz$.
In the spatial phase of \cite{WaleffePF03}, the $(\ell_x, \ell_z)$-periodic 
NBCW solutions have symmetry group 
$\langle \tx \sz, \tz \sxyz \rangle = \{e, \tx \sz, \txz \sxy, \tz \sxyz\}$, 
which is the $S$ symmetry group of \cite{GibsonJFM09} (hereafter GHC09).
\footnote{Note that in \cite{GibsonJFM09}, the $y$ subscript on $\sigma_{xy}$
was suppressed.}
The localizing procedure above sets $[u_g, w_g](x,y,z) = W(z) [u,w](x,y,z)$ 
and determines $v_g$ from incompressibility. A simple series of substitutions 
shows that the first symmetry is preserved under localization, $\bu_g = \tx \sz \bu_g$, 
but the second and third symmetries are not: $\bu_g \neq \txz \sxy \bu_g$ and 
$\bu_g \neq \tz \sxyz \bu_g$. Intuitively, since the windowing function $W(z)$ 
is constant in $x$ and $y$ and even about $z=0$ but not periodic in $z$, 
windowing preserves the $z$-reflection, $x$-translation symmetry 
$\tx \sz$ of the NBCW solution, but not its $\txz \sxy$ or $\tz \sxyz$ symmetries,
which both involve $z$ periodicity. The sole preserved symmetry, $\tx \sz$, is
in fact the symmetry of the localized traveling wave reported in SGB10, 
i.e.\ $[u_g, v_g, w_g](x,y,z) = [u_g, v_g, -w_g](x+\ell_x/2, y, -z)$.
Refinement of this initial guess by a search method that respects symmetry 
results in a traveling-wave solution with the same symmetries. 

The same localization process on a shifted NBWC solution produces an initial
guess with the symmetry of the localized equilibrium solutions of GHC09. 
Shifting the NBCW solution by a quarter-wavelength in $z$, \ie\  
$\tau = \tau(0, \ell_z/4)$, thus changes its symmetry group by conjugation
$\tau s \tau^{-1}$ from 
$\{e, \tx \sz, \txz \sxy, \tz \sxyz\}$ 
to
$\{e, \, \txz \sz, \, \txz \sxy, \, \sxyz\} = \langle \txz \sz, \sxyz \rangle $
which is the $R_{xz}$ symmetry group of GHC09. Of these symmetries, the 
$z$-localization breaks $\txz \sz$ and $\txz \sxy$, since they 
involve periodicity in $z$, and leaves only $\sxyz$ symmetry, which is
in fact the symmetry of the localized equilibrium of plane Couette reported 
in SGB10. 
For choices of $z$ phase that are not integer multiples of $\ell_z/4$, each of 
the three symmetries of the periodic solution is broken by the localization, 
leaving completely unsymmetric initial guesses, corresponding to the rung 
solutions of SGB10.

\subsection{Spanwise localized equilibria of plane Couette flow: computation}
\label{s:spanwisePCF_computation}

\begin{figure}
\begin{tabular}{cc}
{\footnotesize (a)} \hspace{-1.8mm} \includegraphics[width=0.45\textwidth]{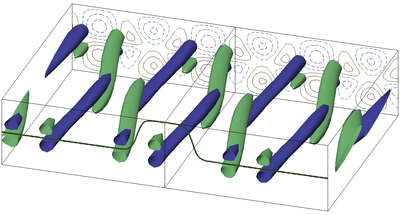} & 
{\footnotesize (b)} \hspace{-1.8mm} \includegraphics[width=0.45\textwidth]{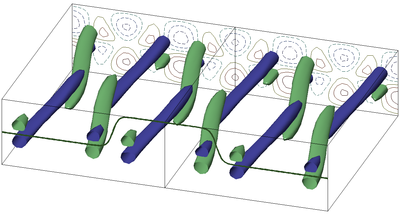} \\
{\footnotesize (c)} \hspace{-1.8mm} \includegraphics[width=0.45\textwidth]{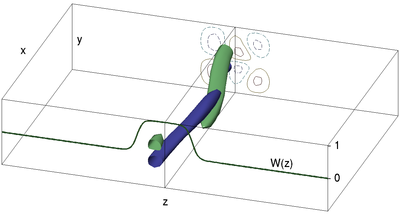} & 
{\footnotesize (d)} \hspace{-1.8mm} \includegraphics[width=0.45\textwidth]{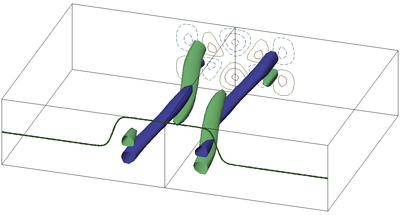} \\
\end{tabular}
\caption{{\bf Construction of localized initial guesses by windowing.} 
(a,b) The spatially periodic EQ7 solution of plane Couette flow, in two spatial phases.
Three copies of the $\alpha, \gamma = 1,2$ periodic solution at $\Rey = 400$ 
are shown in the $z \in [-3\upi/2, \: 3\upi/2]$ subset of the full $L_x, L_z = 2\upi, 6\upi$ computational 
domain, with the $z=0$ plane shown bisecting the box to highlight the $z$ symmetries. 
The solutions in (a) and (b) are related by a quarter-wavelength phase shift in $z$. (c,d) Initial 
guesses for localized solutions with different symmetry groups, produced by multiplying 
(a,b) by windowing functions $W(z)$ (heavy line in front planes). Isosurfaces of signed 
swirling strength at $s = \pm 0.12$ are shown in 
(\colorcomm{green/blue}{light/medium grey}) and contours of streamwise velocity in the 
back plane at contour levels $u = \pm 0.03 \cdot [1\!:\!2\!:\!9]$. 
}
\label{f:guesses} 
\end{figure}

In this section we construct new spanwise localized equilibrium solutions 
of plane Couette flow by applying windowing and refinement
to the spatially periodic EQ7 solution. \refFig{f:guesses} illustrates
how spanwise-localized initial guesses with different symmetry groups are 
constructed from different spatial phases of the spatially periodic solution. 
\refFig{f:guesses}(a) shows EQ7 at $\Rey=400$ with fundamental streamwise and 
spanwise wavenumbers $\alpha,\gamma = 1, 2$ (i.e. periodic lengths 
$\ell_x = 2\upi/\alpha = 2\upi$ and $\ell_z = 2\upi/\gamma = \upi$). The figure shows 
three copies of the periodic structure in the $z \in [3\upi/2, \: 3\upi/2]$ subset 
of the full $L_x,L_z = 2\upi,6\upi$ computational domain. The spatial
phase of EQ7 in \reffig{f:guesses}(a) is chosen so that one concentrated vortex 
structure is centered on the $z=0$ plane. In this phase the solution has 
symmetry group $\langle \sxy, \txz, \tx \sz \rangle$. Each of these symmetries is 
readily apparent in the figure, keeping in mind that the orientation of swirling with 
respect to the $x$ axis and thus \colorcomm{color}{shade} of the isosurfaces changes 
under $\sxy$ (rotation about $z$ axis) and $\tx \sz$ ($x$ shift, $z$-reflect symmetry).
The windowing function $W(z)$ is plotted as a function of $z$ in the front face of the box. 
Multiplication of the periodic structure shown in \reffig{f:guesses}(a) by the windowing 
function, followed by projection onto the divergence-free subspace, produces the initial 
guess shown in \reffig{f:guesses}(c). In this case the windowing parameters $a,b=0.3,1$ 
were chosen to preserve the single concentrated vortex structure centered on the $z=0$ 
plane and to taper rapidly to nearly zero before the next vortical structure. 
As discussed in \refsec{s:windowing},
localization in $z$ breaks the $\txz$ symmetry of the periodic solution, since it 
involves $z$ periodicity, leaving a localized initial guess with symmetry group
$\langle \sxy, \tx \sz \rangle$. Any solution in this symmetry group will be an 
equilibrium, since the $x$ reflection in $\sxy$ prevents travelings waves in $x$
and the $z$ reflection in $\tx \sz$ symmetry prevents traveling waves in $z$.

\refFig{f:guesses}(b,d) illustrate construction of a localized initial guess with
different symmetries by windowing the periodic solution in a different spanwise 
phase. \refFig{f:guesses}(b) shows the same periodic EQ7 solution as in 
\reffig{f:guesses}(a), but translated by a quarter-wavelength in $z$. In this phase 
the periodic NBCW solution has symmetries $\langle \sxy, \sz, \txz \rangle$, which
are again readily apparent in the figure (and which can be derived by conjugating 
$\langle \sxy, \txz, \tx \sz \rangle$ with $\tz^{1/4}$ and choosing the specified 
symmetries as generators for the conjugated group). A wider windowing function, with
$a = b = 1$, preserved a pair of mirror-symmetric concentrated vortical structures in 
the core region and tapered rapidly to nearly zero before reaching the next vortices,
as shown in \reffig{f:guesses}(b). The windowing breaks symmetries with factors of $\txz$, 
producing an initial guess for a spanwise-localized solution with symmetry group 
$\langle \sxy, \sz \rangle$. Again, sign changes in all three coordinates in this 
symmetry group fix the phase of the velocity field with respect to the origin and rule 
out traveling waves. Thus, the localized solutions in {\em both} choices of $z$ phase 
will be equilibria, unlike the localized solutions of SGB10, where one choice of phase 
produces equilibria and the other streamwise-traveling waves.

The localized initial guesses depicted in \reffig{f:guesses}(c,d) were then 
refined to numerically exact equilibrium solutions of plane Couette flow shown in
\reffig{f:spanwise_couette}(a,b), using a Newton-Krylov-hookstep search algorithm.
The search algorithm finds equilibria as 
solutions of the equation $\bff^T(\bu) - \bu = 0$, where $\bff^T$ is the time-$T$ 
integration of the discretized Navier-Stokes equations with appropriate boundary 
conditions. The time integration is performed with a Fourier-Chebyshev-tau scheme 
in primitive variables (\cite{SpalartJCP91,Canuto06}) and 3rd-order semi-implicit 
backwards differentiation time stepping (\cite{Peyret02}). The Fourier-Chebyshev 
spatial discretization of $\bu(\bx,t)$ takes the form
\begin{equation}
\bu(\bx) = \sum_{j=-J}^J \sum_{k=-K}^K \sum_{\ell=0}^L \hat{\bu}_{jkl} T_{\ell}(y) e^{2\upi i (jx/L_x + kz/Lz)}
\end{equation}
Spatial discretization levels are specified by $(J,K,L)$ or by the corresponding 
$N_x \times N_y \times N_z = (2J\!+\!1) \times L \times (2K\!+\!1)$ physical grid.
Nonlinear terms are computed with 3/2-style dealiasing. We set spatial discretization 
levels so the maximum truncated Fourier and Chebyshev modes are $O(10^{-6})$ and 
$O(10^{-10})$ respectively. Experience has taught us that coarser spatial discretizations 
can result in spurious solutions. Symmetries were enforced through the search by 
projecting $\bu \rightarrow (\bu + \sigma \bu)/2$ for each of the generators $\sigma$ of 
the appropriate symmetry group at the intervals $\Delta T =1$ during time integration. 
The residual of the search equation is $\| \bff^T(\bu) - \bu \|/T$ using the $L^2$ 
norm 
\begin{align}
 \| \bu \| &= \left[\frac{1}{V} \int_V \bu \cdot \bu \, d\bx\right]^{1/2}
\label{L2norm}
\end{align}
where $V$ is the volume of the computational domain. We measure the accuracy of a 
given solution by increasing spatial resolution by a factor of 3/2 in each direction, 
decreasing the time step by a factor of 2, and then recomputing the residual. Further 
details of the implementation of the search algorithm and time integration are given 
in GHC09, and the code is available for download at {\tt www.channelflow.org} (\cite{chflow}). 

The localized plane Couette solutions depicted in \reffig{f:spanwise_couette}(a,b)
were computed at $\Rey=400$ in a $2\upi,6\upi$ computational box and a 
$16 \times 65 \times 128$ grid, with integration time $T=10$ and time step 
$dt=0.077$, resulting in a CFL number of about 0.6. The residuals of the windowed 
initial guesses began at roughly $10^{-3}$ and were reduced to $10^{-15}$ after 
six or seven iterations of the Newton-Krylov-hookstep algorithm. The  $O(10^{-6})$
accuracy of these solutions, measured as discussed above, is the best one could 
expect given that spectral coefficients were truncated at that level. The tails of 
the localized solutions drop to $O(10^{-4})$ at the $z=\pm 3\upi$ edge of the 
computational domain. The computational cost is modest: about one CPU-hour for each 
solution, running serially on a desktop computer with a 3.3 GHz Intel i7-3960X processor.

\subsection{Spanwise localized equilibria of plane Couette flow: roll-streak structure}
\label{s:spanwisePCF_structure}

\begin{figure}
\begin{tabular}{cc}
{\footnotesize (a)} \hspace{-1.8mm} \includegraphics[width=0.45\textwidth]{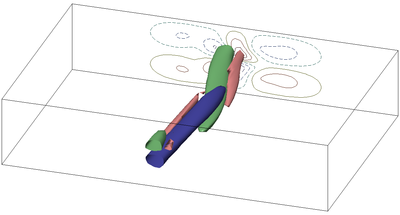} & 
{\footnotesize (b)} \hspace{-1.8mm} \includegraphics[width=0.45\textwidth]{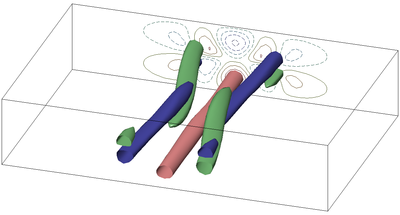} \\
{\footnotesize (c)} \hspace{-1.8mm} \includegraphics[width=0.45\textwidth]{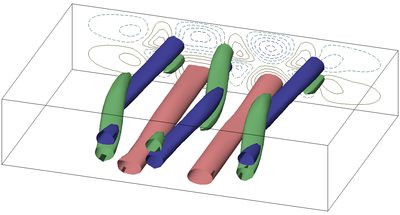} &
{\footnotesize (d)} \hspace{-1.8mm} \includegraphics[width=0.45\textwidth]{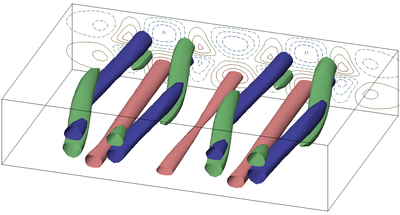} \\ 
{\footnotesize (e)} \hspace{-1.8mm} \includegraphics[width=0.45\textwidth]{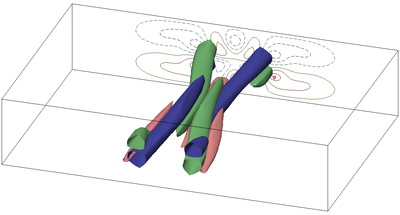} &
{\footnotesize (f)} \hspace{-1.8mm} \includegraphics[width=0.45\textwidth]{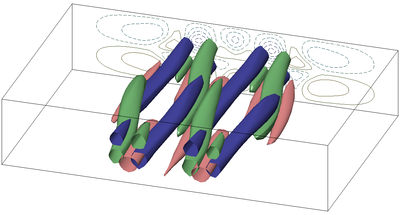} 
\end{tabular}
\caption{{\bf Spanwise-localized equilibrium solutions of plane Couette flow} 
constructed from windowing the spatially periodic EQ7 equilibrium of 
\cite{GibsonJFM09} at $\Rey = 400$, in the $z \in [-3\upi/2, 3\upi/2]$ subset of 
the $L_x, L_z = 2\upi, 6\upi$ computational domain. (a,b) are localized solutions 
produced from the initial guesses shown in \reffig{f:guesses}(c,d). See text for 
discussion of (c-f). The plotting conventions are the same as in \reffig{f:guesses}, 
but additionally isosurfaces of streamwise velocity at $u = 0.18$ 
are plotted in \colorcomm{red}{dark gray} to show the positions of high-speed 
streaks near the lower wall (symmetric high-speed streaks near the upper wall 
are not shown).}
\label{f:spanwise_couette}
\end{figure}

\refFig{f:spanwise_couette} shows six spanwise-localized equilibrium solutions 
of plane Couette flow with streamwise wavenumber $\alpha = 1$ and $\Rey=400$. 
\refFig{f:spanwise_couette}(a-d) were obtained by the localization and search 
methods outlined in \refsec{s:spanwisePCF_computation}, (a,b) as described 
in detail in \refsec{s:spanwisePCF_computation} and (c,d) by increasing the 
core region of the window to fit three and four copies of the concentrated 
vortex structures respectively. 
Solutions (a-d) show that it is possible to obtain localized versions
of EQ7 with 1,2,3, and 4 copies of the basic concentrated vortical structure 
shown in isolation in (a), by choosing appropriate centers of symmetry and 
width of the core region of the windowing function. We will refer to the
solutions (a,b,c,d) as EQ7-1,2,3,4 respectively. Odd-numbered solutions 
have $\langle \sxy, \tx \sz \rangle$ symmetry and even-numbered solutions 
$\langle \sxy, \sz \rangle$.  We note that each of the
EQ7-1,2,3,4 solutions lies on a distinct solution branch. This is 
in distinct contrast to the localized versions of the NBCW solutions, for which 
all solutions with the same symmetry lie on a single solution branch, and 
additional copies of the fundamental structure are added in a continuous 
fashion under continuation in Reynolds number, via homoclinic snaking
(\cite{SchneiderPRL10}). The localized EQ7 solutions, in contrast, can each 
be continued in Reynolds number around a single saddle-node bifurcation,
and solutions on the opposite branch have distinctly different physical 
structure. For example, \reffig{f:spanwise_couette}(e) and (f) show the
opposite branches of (b) EQ7-2 and (d) EQ7-4, obtained by continuing these 
downward in Reynolds, around a saddle-node bifurcation, and back upwards to 
$\Rey=400$.

\begin{figure}
\begin{tabular}{cc}
{\footnotesize (a)} \hspace{-1.8mm} \includegraphics[width=0.45\textwidth]{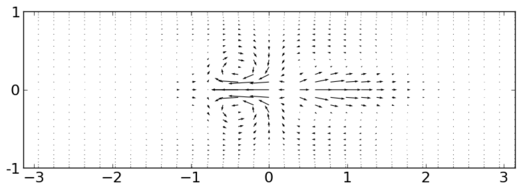} & 
{\footnotesize (f)} \hspace{-1.8mm} \includegraphics[width=0.45\textwidth]{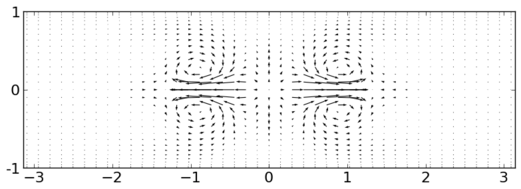} \\
{\footnotesize (b)} \hspace{-1.8mm} \includegraphics[width=0.45\textwidth]{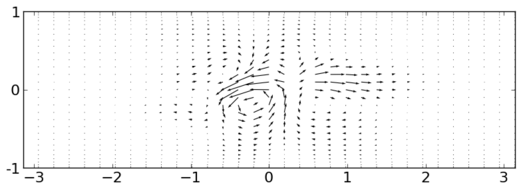} & 
{\footnotesize (g)} \hspace{-1.8mm} \includegraphics[width=0.45\textwidth]{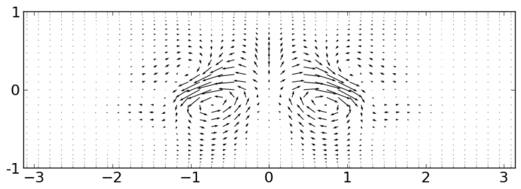} \\
{\footnotesize (c)} \hspace{-1.8mm} \includegraphics[width=0.45\textwidth]{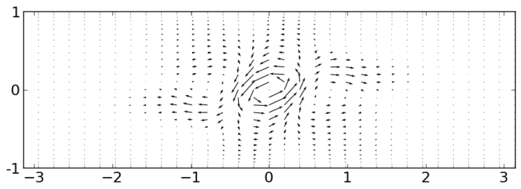} & 
{\footnotesize (h)} \hspace{-1.8mm} \includegraphics[width=0.45\textwidth]{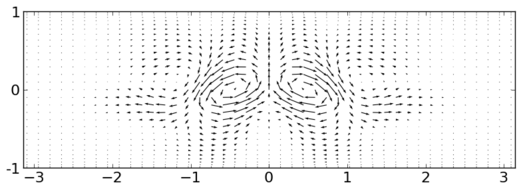} \\
{\footnotesize (d)} \hspace{-1.8mm} \includegraphics[width=0.45\textwidth]{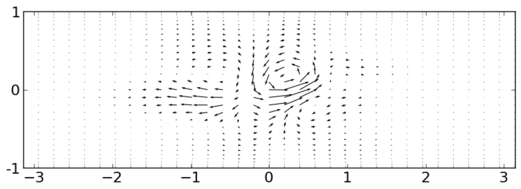} & 
{\footnotesize (i)} \hspace{-1.8mm} \includegraphics[width=0.45\textwidth]{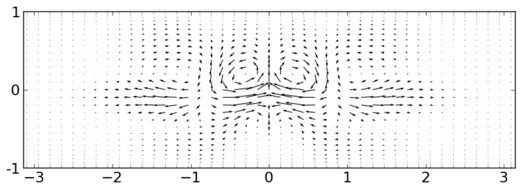} \\
{\footnotesize (e)} \hspace{-1.8mm} \includegraphics[width=0.45\textwidth]{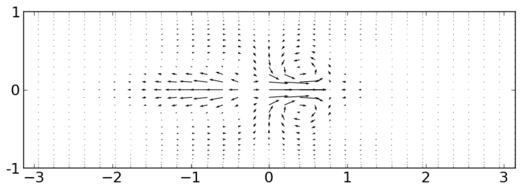} & 
{\footnotesize (j)} \hspace{-1.8mm} \includegraphics[width=0.45\textwidth]{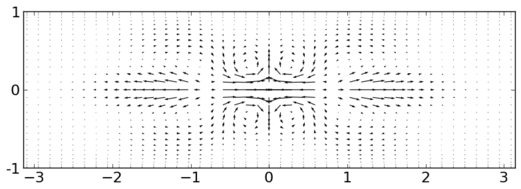} 
\end{tabular}
\caption{{\bf Cross-sections of spanwise-localized equilibrium solutions of plane Couette flow.}
(a-e) Quiver plots of  $[v,w](y,z)$ for EQ7-1 at $x = \{-\upi, -3\upi/4, -\upi/2, -\upi/4, 0\}$
with $y$ vertical and $z$ horizontal.  
(f-j) EQ7-2 at same $x$ values. $\alpha =1$ and $\Rey=400$ for both. Compare to the same solutions 
shown with isosurfaces of signed swirling strength in \reffig{f:spanwise_couette}(a) and (b); the 
cross-sections here are evenly spaced along the front half of those figures. The $z \in [-\upi,\upi]$ 
subset of the full $z \in [-3\upi,3\upi]$ computational domain is shown. 
}
\label{f:couette_crosssections}
\end{figure}

The spatial structure of EQ7-1 and EQ7-2 is illustrated in more detail in 
\refFig{f:couette_crosssections}. \refFig{f:couette_crosssections}(a-e) 
show the cross-stream velocity $[v,w](yz)$ for EQ7-1 in five streamwise-normal cross-sections 
spaced evenly between $x=-\upi$ and $x=0$, which are the front face and the middle of the boxes
in \reffig{f:spanwise_couette}(a,b). The \colorcomm{blue}{medium gray} isosurface of signed 
swirling strength in the front half of the box in \reffig{f:spanwise_couette}(a) 
appears here as a concentrated counter-clockwise vortex that begins just below and to the left of 
the origin at $x=-\upi$ in (a), increases in strength and moves upward and to the right in (b-d),
and ends above and to the right of the origin at $x=0$ in (e). By the $\tx \sz$ symmetry of EQ7-1, 
the equivalent quiver plots for $x=0$ through $x=\pi$ would be the $z$-mirror images of (a-e),
showing a concentrated clockwise vortex starting below and to the right of the origin and 
moving upwards and leftwards, and corresponding to the \colorcomm{green}{light gray} isosurface 
of signed swirling strength in the back half of the box in \reffig{f:spanwise_couette}(a). 
Likewise, the predominant features of EQ7-2 shown in \reffig{f:spanwise_couette}(f-j) are two 
$z$-symmetric counter-rotating vortices that begin near the lower wall at $x=-\upi$ in (f),
and rise upwards and closer together in (g-j). Due to EQ7-2's $\sxy$ symmetry, cross-sections 
from $x=0$ to $\upi$ would appear as the $y$-mirror images of (j-f), with a pair of nearby 
counter-rotating vortices in the lower half of the $(y,z)$ plane rising and separating, as in 
the back half of \reffig{f:spanwise_couette}(b). The clear and distinct concentrations 
of circulating fluid in \reffig{f:couette_crosssections} constitute our main justification for 
speaking of ``regions of concentrated vortical structure'' and show that isosurface plots of 
signed swirling strength are not misleading but rather show precisely where such regions lie.

\begin{figure}
\begin{tabular}{cc}
{\footnotesize (a)} \hspace{-1.8mm} \includegraphics[width=0.45\textwidth]{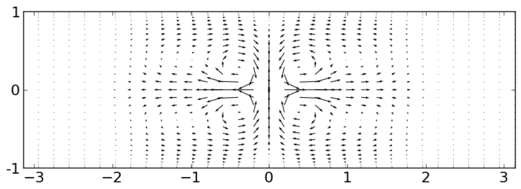} & 
{\footnotesize (c)} \hspace{-1.8mm} \includegraphics[width=0.45\textwidth]{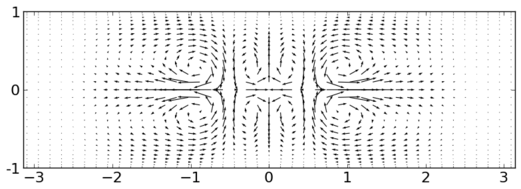} \\
{\footnotesize (b)} \hspace{-1.8mm} \includegraphics[width=0.45\textwidth]{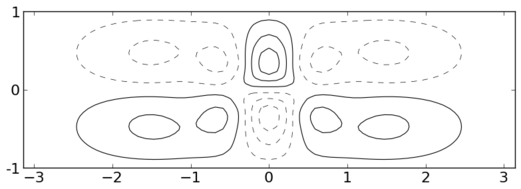} & 
{\footnotesize (d)} \hspace{-1.8mm} \includegraphics[width=0.45\textwidth]{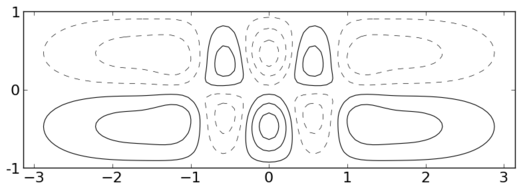} 
\end{tabular}
\caption{{\bf Streamwise-averaged roll-streak structure of spanwise-localized equilibria of 
plane Couette flow.} 
(a) Quiver plot of $x$-average $[v,w](y,z)$ and (b) contour plot of $x$-average $u(y,z)$ for
EQ7-1 with $y$ vertical and $z$ horizontal. (c,d) Same for EQ7-2. Both solutions are shown at 
$\alpha = 1$ and $\Rey=400$. 
Contour lines are plotted at levels $u= \pm\{0.03, 0.09, 0.15\}$, with negative values in 
dashed lines positive in solid. Quiver plots are autoscaled. The $z \in [-\upi,\upi]$ subset 
of the full $z \in [-3\upi,3\upi]$ computational domain is shown. 
}
\label{f:couette_xaverages}
\end{figure}

The mean roll-streak structure of EQ7-1 and EQ7-2 is illustrated in \refFig{f:couette_xaverages}. 
The four counter-rotating vortices surrounding the origin in \reffig{f:couette_xaverages}(a) 
result from $x$-averaging the counter-clockwise vortex that slopes upwards and rightwards 
from $x=-\upi$ to $0$ in \reffig{f:spanwise_couette}(a) and \reffig{f:couette_crosssections}(a-e)
with its clockwise $\tx \sz$-symmetric counterpart that slopes upwards and leftwards from $x=0$ 
to $\upi$. These four vortices create the pattern of alternating positive and negative streamwise 
streaks (relative to laminar flow) shown in (b) by advecting high-speed fluid ($u_{\text{tot}} = \pm1$) 
from the walls towards the interior in the region near $z =0$, and low-speed fluid 
($u_{\text{tot}} \approx 0$) from interior towards the walls for larger $z$. 
\refFig{f:spanwise_couette}(c,d) shows the corresponding mean roll-streak structure for 
EQ7-2; here the doubling of the basic concentrated vortex structure compared to EQ7-1,
apparent in \reffig{f:spanwise_couette}(b), results in the eight counter-rotating mean 
vortices shown in (c) and an increased pattern of alternating streamwise streaks shown in 
(d).

\section{Traveling wave solutions of channel flow}
\label{s:channelsolns}




In this section we extend the results of \refsec{s:planecouette} to channel-flow 
conditions.
For the first set of solutions, we use a numerical continuation method  
similar in spirit to Waleffe's homotopy of the NBCW solution between plane 
Couette and ``half-Poiseuille'' flow with no-slip at the lower wall and 
free-slip conditions at the upper wall (\cite{WaleffePRL98, WaleffeJFM01}). 
When extended by symmetry to full channel conditions, Waleffe's continuation 
produces a traveling-wave solution symmetric about the channel midplane, with 
two NBCW-like roll-streak structures, each positioned in the high-shear region 
near either wall, and mirror symmetric ($\sy$) to each other across the midplane. 

In the present study, we continue the EQ7 solutions from plane Couette to full 
channel conditions, enforcing no-slip boundary conditions on both walls throughout.
The continuation is done in two stages: first with fixed wall speed and increasing 
pressure gradient, then fixed pressure gradient and decreasing wall speed.  
As in \refsec{s:eqns_PCF}, we decompose the total velocity field into a base 
flow and a deviation, $\butot(\bx,t) = \bu(\bx,t) + U(y) \, \be_x$, and the 
total pressure field into $\ptot = p(\bx,t) + x \; \dPdx$, where $dP/dx$ is
a parametric constant corresponding to the externally imposed mean pressure 
gradient. To make the decomposition unique, we specify that the fluctuation 
pressure $p$ is periodic (so that the spatial mean of $\grad p$ is zero), and 
that the laminar velocity profile satisfies the no-slip conditions at the walls 
and balances the imposed mean pressure gradient, $\dPdx = \nu U''$, where $\nu$ is the 
kinematic viscosity of the fluid. Consequently the base flow is the laminar 
solution for the given viscosity, mean pressure gradient, and wall speed, and 
the fluctuation velocity satisfies Dirichlet conditions at the walls. 
The Navier-Stokes equations again take the form of \refeq{eq:NSE_PCF}.
The Reynolds number $\Rey=\bar{U}h/\nu$ is based on a velocity scale $\bar{U}$ 
appropriate to the flow as it transforms from plane Couette to pressure-driven 
channel conditions, namely, $\bar{U}$ is half the relative wall speed when 
continuing in pressure gradient and the centerline velocity of the laminar 
base flow when continuing in wall speed ($\bar{U} = |\dPdx|\; h^2/(2\nu)$).
Thus in nondimensional terms the continuation is first in mean pressure gradient 
$dP/dx$ from 0 to $-2/\Rey$ with wall speeds fixed at $U(\pm 1) = \pm 1$ (equivalently 
from $U(y) = y$ to $U(y) = 1 + y - y^2$), and then continuation in wall speed from 
1 to 0 with mean pressure gradient held fixed at $P_x = -2/\Rey$ (equivalently from 
$U(y) = 1 + y - y^2$ to $U(y) = 1-y^2$). For channel flow conditions $U(y) = 1 - y^2$ 
and $\dPdx \neq 0$, eqn.\ \refeq{eq:NSE_PCF} and boundary conditions are 
invariant under any combination of $x$ and $z$ translations and reflections about 
the $y\!=\!0$ and $z\!=\!0$ midplanes; thus the symmetry group of channel flow is 
$\langle \sy, \sz, \tau(\Delta x, \Delta z) \rangle$, where 
$\sy : [u,v,w](x,y,z) \rightarrow [u,-v, w](x,-y, z)$.

\subsection{Spatially periodic traveling wave solutions of channel flow}
\label{s:periodic_channel}

\begin{figure}
\begin{center}
\begin{tabular}{ccc}
\hspace{-0mm} {\footnotesize (a)} \hspace{-2mm} \includegraphics[width=0.30\textwidth]{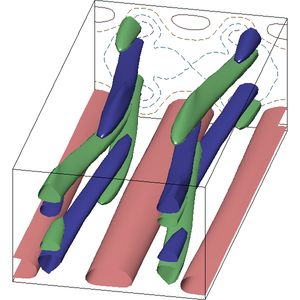} & 
\hspace{-4mm} {\footnotesize (b)} \hspace{-2mm} \includegraphics[width=0.30\textwidth]{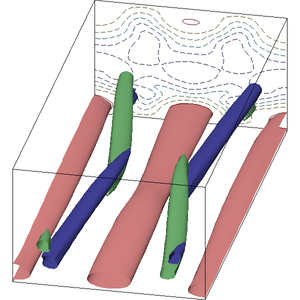} & 
\hspace{-4mm} {\footnotesize (c)} \hspace{-2mm} \includegraphics[width=0.30\textwidth]{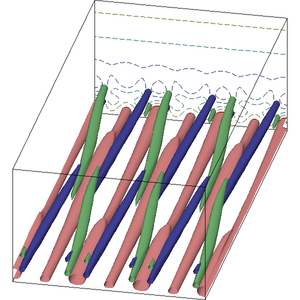} 
\end{tabular}
\end{center}
\caption{{\bf Spatially periodic traveling waves of channel flow with decreasing wall-normal symmetry.}
(a) TW1, constructed by continuing EQ7 from plane Couette to channel conditions, has symmetry
group $\langle  \sy, \sz, \txz \rangle$. Isosurfaces of signed swirling strength are at 
$s = \pm 0.04$ are shown in \colorcomm{green/blue}{light/medium grey}. High-speed streaks 
near the lower wall are shown by isosurfaces of streamwise velocity at $u=0.02$. High-speed 
streaks near the upper wall, symmetric to those near the lower wall, are not shown as 
isosurfaces but they are indicated in the contour plots of streamwise velocity in the back 
plane. 
(b) TW2, found from an initial guess judiciously chosen from numerical simulation data, has symmetry 
group $\langle \sz, \txz \rangle$. Isosurfaces of signed swirling strength at $s = \pm 0.10$ are 
shown in \colorcomm{green/blue}{light/medium grey}, and high-speed streaks near the lower wall 
are shown by isosurfaces of streamwise velocity at $u=0.03$. There are similar but weaker 
vortex structures and high-speed streaks near the upper wall, but they do not appear at 
these levels for the isosurfaces. 
Both (a) and (b) are shown at $\alpha, \gamma = 1, 2$ and $\Rey = 2300$. 
(c) TW3, constructed by continuation in Reynolds number from TW2 and further symmetrization 
and Newton-Krylov refinement, has $\langle \sz, \txz \rangle$ but periodicity $\alpha, \gamma = 1, 6$,
and is shown at $\Rey=4000$. Isosurfaces of signed swirling strength are at $s = \pm 0.9$ and
$u=0.03$. Each back plane shows contour plot of streamwise velocity at levels 
$u = \{0.03, -0.03, -0.09, -0.15, ...\}$, positive in solid lines and negative in dashed.
}
\label{f:periodic_channel_swirling}
\end{figure}

\begin{figure}
\begin{tabular}{ccc}
{\footnotesize (a)} \hspace{-2.4mm} \includegraphics[width=0.29\textwidth]{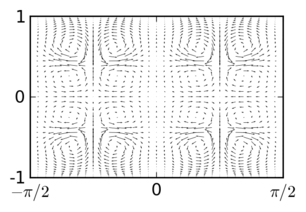} &
{\footnotesize (b)} \hspace{-2.4mm} \includegraphics[width=0.29\textwidth]{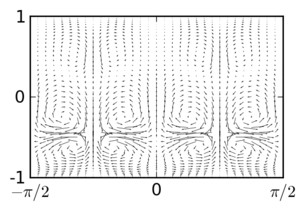} &
{\footnotesize (c)} \hspace{-2.4mm} \includegraphics[width=0.29\textwidth]{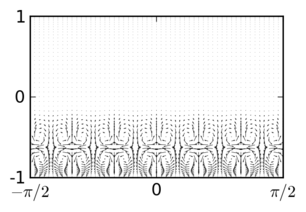} \\
{\footnotesize (d)} \hspace{-2.4mm} \includegraphics[width=0.29\textwidth]{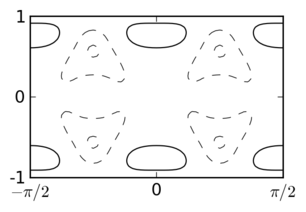} &
{\footnotesize (e)} \hspace{-2.4mm} \includegraphics[width=0.29\textwidth]{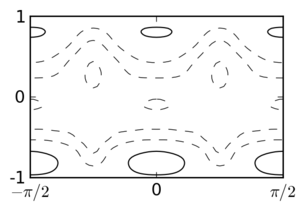} &
{\footnotesize (f)} \hspace{-2.4mm} \includegraphics[width=0.29\textwidth]{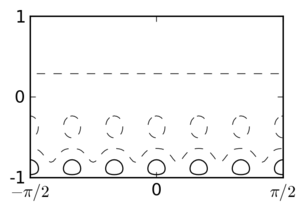} \\
\end{tabular}
\caption{{\bf Streamwise-averaged roll-streak structure of spanwise-periodic traveling waves
of channel flow.}
Quiver plots of $x$-average $[v,w](y,z)$ and contour plots of $x$-average $u(y,z)$ for 
(a,d) TW1, (b,e) TW2, and (c,f) TW3, with the same spatial and Reynolds parameters
as in \reffig{f:periodic_channel_swirling}. Contour lines are shown at levels $u = {-0.2, -0.1}$ 
(dashed) and $u=0.025$ (solid), with an additional negative contour line is shown at $u= -0.3$ 
in (e). The quiver plots are autoscaled. Axes are $z$ horizontal and $y$ vertical.
}
\label{f:periodic_channel_xaverage}
\end{figure}

{\bf TW1:} 
\refFig{f:periodic_channel_swirling}(a) and \reffig{f:periodic_channel_xaverage}(a,d) 
show a spatially periodic traveling-wave solution of channel flow with symmetry 
$\langle \sy, \sz, \txz \rangle$ obtained by continuation from plane Couette conditions, 
as described above. The starting point for continuation was the spatially periodic EQ7 
equilibrium of plane Couette flow at $\Rey=2000$ and $\alpha,\gamma =1,2$; the
TW1 channel traveling wave has the same spatial periodicity and is shown at $\Rey=2300$. 
Note that TW1 is mirror symmetric about the $y=0$ midplane. The $y$ mirror symmetry is 
most clearly seen in the streamwise-averaged plots of cross-stream and streamwise velocity 
in \reffig{f:periodic_channel_xaverage}(a,d). Although $\sy$ symmetry is within the 
symmetry group of channel flow, we did not expect the continuation to produce a solution 
with this symmetry, since it was neither present in the initial plane Couette solutions 
nor allowed in the intermediate steps in the continuation from plane Couette to channel 
conditions. Instead, we expected that the increasing $y$ asymmetry under continuation 
in pressure gradient would push the vortex structures towards the lower wall, where the
shear of the base flow is higher ($|U'(-1)| = 3$, compared to $|U'(1)| = 1$, for the 
base flow $U(y) = 1 + y - y^2$ attained at the end of the pressure 
continuation), and that this $y$-asymmetry would be maintained during continuation 
in wall speed down to $U(y) = 1 - y^2$. However, it turned out that weak vortices 
formed near the upper wall under pressure continuation and grew in strength during 
wall-speed continuation so that the solution gained $\sy$ symmetry when the wall 
speed reached zero. In terms of the symmetry groups, the starting plane Couette 
solution had symmetry group $\langle \sxy, \sz, \txz \rangle$, continuation to nonzero 
$dP/dx$ broke $\sxy$ symmetry, and $\sy$ was gained at the final step of wall-speed 
continuation, resulting in symmetry $\langle \sy, \sz, \txz \rangle$.
The structure of vortices and streaks in TW1 can roughly be described as two copies
of EQ7 stacked on top of each other, with the upper copy either phase-shifted 
by half a wavelength in $x$ or having $[v,w]$ reversed in sign via mirror symmetry 
in $y$. This is apparent from comparison of TW1 in \reffig{f:periodic_channel_swirling}(a) 
to EQ7 in \reffig{f:EQ1_EQ7}(b) and EQ7-1 in \ref{f:spanwise_couette}(b).

{\bf TW2:} \refFig{f:periodic_channel_swirling}(b) and \reffig{f:periodic_channel_xaverage}(b,e) 
show a spatially periodic traveling-wave solution of channel flow with asymmetry in $y$ and
symmetry $\langle \sz, \txz\rangle$, obtained from an initial guess from turbulent simulation 
data (\cite{ViswanathJFM07,GibsonJFM09}). 
Specifically, we $z$-mirror-symmetrized an arbitrary turbulent velocity field of channel flow 
at $\Rey=3750$ in a $2\upi,1\upi$ box by applying $\bu \rightarrow 1/2 \; (1 + \sz) \bu$ 
and then quenched the turbulent field by lowering the Reynolds number and 
continuing time integration with the bulk velocity fixed at 2/3 and the $\sz$ 
symmetry enforced by projection at regular intervals. After some experimentation,
we found that after quenching to $\Rey=2650$, the fine-scale structure 
of the velocity field and the spatial-mean wall shear decreased quickly, the 
latter reaching a local minimum after about 50 time units and growing slowly 
again for another 50 time units before resuming a high level of wall shear 
with rapid fluctuations. The smoothness and length of this minimum suggested 
a close pass to a hyperbolic edge state. Using a velocity field from this
minimum as an initial guess for a Newton-Krylov search produced a numerically 
exact spatially periodic, wall-localized traveling wave solution of channel
flow with $\langle \sz, \txz\rangle$ symmetry.

The $y$ asymmetry of TW2 is exaggerated in \reffig{f:periodic_channel_swirling}(b) by the 
binary character of isosurface plots. In fact at $\Rey=2300$ TW2 has weaker vortex structures 
near the upper wall with swirling strength comparable to those of TW1, and weaker streaks 
there as well. Both the vortices and streaks near the upper wall are visible in the 
streamwise-average plots of TW2 in \reffig{f:periodic_channel_xaverage}(b,e). Note also 
that the swirling strength isosurfaces of TW2 in \reffig{f:periodic_channel_swirling}(b) 
are at $s = \pm 0.10$, over twice the magnitude of those for TW1 in 
\reffig{f:periodic_channel_swirling}(a) at $s = \pm 0.4$. 
Compared to the four layers of counter-rotating mean vortices stacked symmetrically 
about $y=0$ in TW1 (see \refFig{f:periodic_channel_xaverage}(a,d)), TW2 has two layers 
of counter-rotating mean vortices below $y=0$ and a single layer of vortices above $y=0$,
and these vortices have the same orientation as the vortices below them. And though 
it is not clear from the autoscaled quiver plots, the mean vortices of TW2 near the 
lower wall are about three times the magnitude of those of TW1, as measured by 
magnitudes of the $[v,w]$ velocities, and the mean vortices of TW2 above $y=0$ are 
of comparable magnitude to those of TW1. However, the $y$-asymmetry of TW2 increases
as the Reynolds number is increased (see TW3) and as $z$-periodicity is relaxed (see
TW2-1 and TW2-2). 

{\bf TW3:} \refFig{f:periodic_channel_swirling}(c) and \reffig{f:periodic_channel_xaverage}(c,f) 
show a wall-localized, spanwise and streamwise periodic traveling wave of channelflow
with symmetry $\langle \sz, \txz\rangle$, discovered through continuation of TW2 in Reynolds 
number. The fundamental $z$ wavenumber of TW2 is $\gamma = 2$, but as as $\Rey$ increased towards 
$4000$, the structure at this wavenumber weakened and structure at $\gamma = 6$ grew, while the 
structure away from the lower wall weakened substantially until it became nearly laminar. TW3 
was computed by zeroing all modes in TW2 at $\Rey=4000$ with $\gamma < 6$ and refining this initial 
guess to an exact traveling wave with Newton-Krylov-hookstep search. The resulting TW3 solution 
has periodicity $\alpha,\gamma = 1,6$ and symmetry $\langle \sz, \txz\rangle$, where $\txz$
is understood as involving a half-cell shift in $z$ with respect to the smaller $\ell_z=\upi/3$
periodic length. Its most notable property is its very strong localization in the wall-normal 
direction, as evidenced by \reffig{f:periodic_channel_xaverage}(c,f).

\subsection{Spanwise localized traveling wave solutions of channel flow}
\label{s:spanwise_channel}

\begin{figure}
\begin{tabular}{cc}
{\footnotesize (a)} \hspace{-1.8mm} \includegraphics[width=0.45\textwidth]{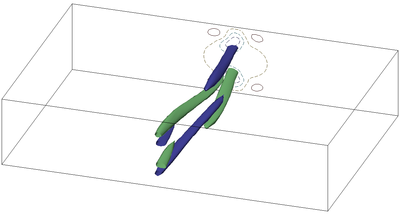} & 
{\footnotesize (b)} \hspace{-1.8mm} \includegraphics[width=0.45\textwidth]{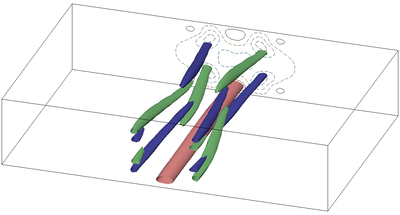} \\ 
{\footnotesize (c)} \hspace{-1.8mm} \includegraphics[width=0.45\textwidth]{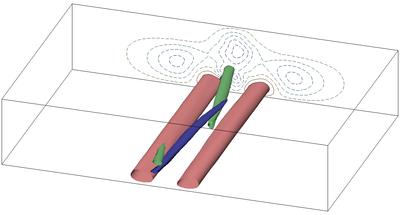} &
{\footnotesize (d)} \hspace{-1.8mm} \includegraphics[width=0.45\textwidth]{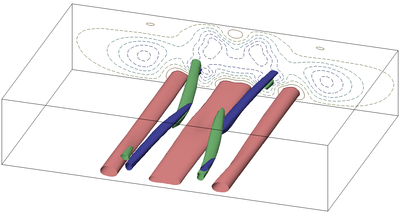} \\
\end{tabular}
\caption{{\bf Spanwise-localized traveling wave solutions of channel flow.}
(a) TW1-1, (b) TW1-2, (c) TW2-1, and (d) TW2-2. 
(a,b) TW1-1 and TW1-2 are  $\langle \sy, \tx \sz\rangle$ and $\langle \sy, \sz\rangle$ 
symmetric traveling waves obtained by localizing TW1 in two different $z$ phases.
(c,d) TW2-1 and TW2-2 are $\tx \sz$ and $\sz$ symmetric traveling waves obtained by 
localizing TW2 in two different $z$ phases. Plotting conventions are the same as in 
\reffig{f:periodic_channel_swirling}, but with isosurfaces of signed swirling strength 
and streamwise velocity at (a,b) $s = \pm 0.05$, $u=0.05$, and (c,d) $s = \pm 0.09$, 
$u=0.08$. Contour lines of streamwise velocity are shown in the back $y,z$ plane at 
levels $u=0.03$ (solid lines) and (a,b) $u = \{-0.03, -0.09, -0.15\}$, (c,d) 
$u = \{-0.03, -0.09, -0.15, -0.21, -0.30\}$ (dashed). In (b) a high-speed streak near 
the upper wall, symmetric to that near the lower wall, is suppressed to avoid visual 
clutter. By contrast, (c,d) show true asymmetry in high-speed speed streaks:
(c) has high-speed streaks near the lower wall only, and in (d) the streaks near the 
upper wall are substantially weaker, below the given isosurface levels. Solutions are 
shown at $\Rey=2300$ in $z \in [-3\upi/2, \: 3\upi/2]$ subsets of their full 
$L_x,L_z=2\upi,6\upi$ computational domains.
}
\label{f:spanwise_channel}
\end{figure}

\begin{figure}
\begin{tabular}{cc}
{\footnotesize (a)} \hspace{-1.8mm} \includegraphics[width=0.45\textwidth]{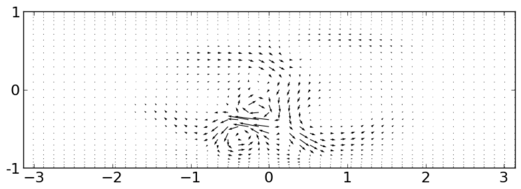} & 
{\footnotesize (f)} \hspace{-1.8mm} \includegraphics[width=0.45\textwidth]{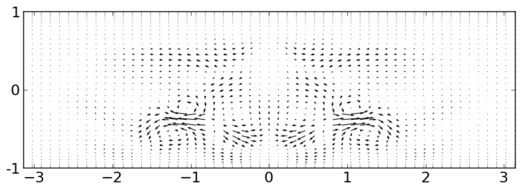} \\
{\footnotesize (b)} \hspace{-1.8mm} \includegraphics[width=0.45\textwidth]{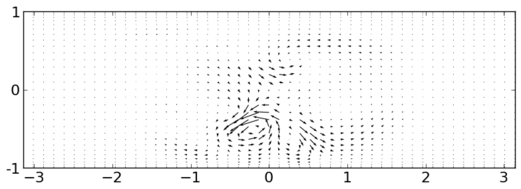} & 
{\footnotesize (g)} \hspace{-1.8mm} \includegraphics[width=0.45\textwidth]{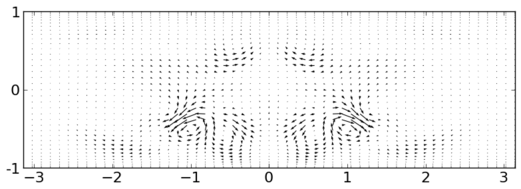} \\
{\footnotesize (c)} \hspace{-1.8mm} \includegraphics[width=0.45\textwidth]{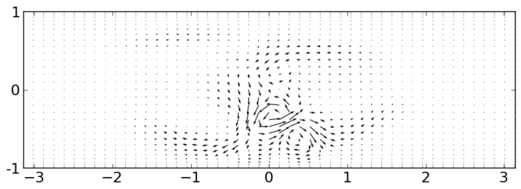} &
{\footnotesize (h)} \hspace{-1.8mm} \includegraphics[width=0.45\textwidth]{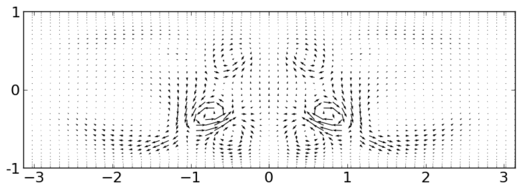} \\
{\footnotesize (d)} \hspace{-1.8mm} \includegraphics[width=0.45\textwidth]{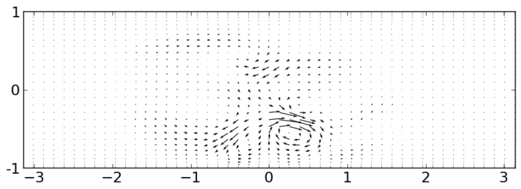} & 
{\footnotesize (i)} \hspace{-1.8mm} \includegraphics[width=0.45\textwidth]{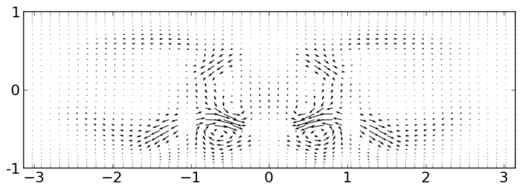} \\
{\footnotesize (e)} \hspace{-1.8mm} \includegraphics[width=0.45\textwidth]{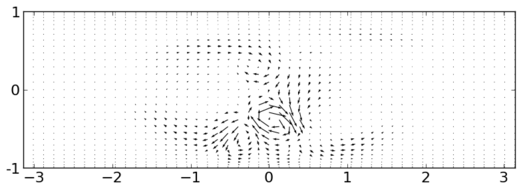} & 
{\footnotesize (j)} \hspace{-1.8mm} \includegraphics[width=0.45\textwidth]{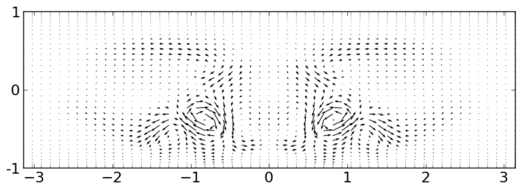} 
\end{tabular}
\caption{{\bf Cross-sections of spanwise-localized, near-wall traveling waves of 
channel flow.}
(a)-(e) Quiver plots of $[v,w](y,z)$ for TW2-1 at $x = \{-\pi, -0.6 \pi, -0.2\pi, 0.2\pi, 0.6\pi\}$, 
(f)-(j) same for TW2-2, with $y$ vertical and $z$ horizontal. Both solutions are at $\alpha=1$ and 
$\Rey=2300$.  The $z \in [-\upi,\upi]$ subset of the full $z \in [-3\upi,3\upi]$ computational domain
is shown.
}
\label{f:spanwise_channel_xsect}
\end{figure}

\begin{figure}
\begin{tabular}{cc}
{\footnotesize (a)} \hspace{-1.8mm} \includegraphics[width=0.45\textwidth]{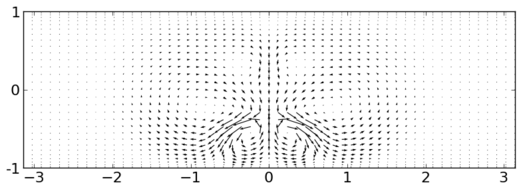} & 
{\footnotesize (c)} \hspace{-1.8mm} \includegraphics[width=0.45\textwidth]{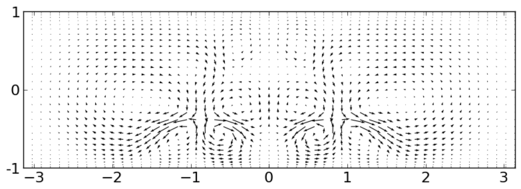} \\
{\footnotesize (b)} \hspace{-1.8mm} \includegraphics[width=0.45\textwidth]{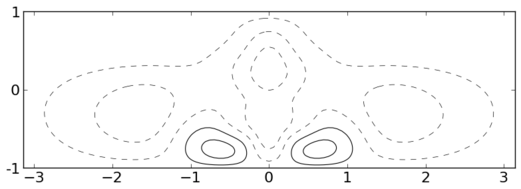} & 
{\footnotesize (d)} \hspace{-1.8mm} \includegraphics[width=0.45\textwidth]{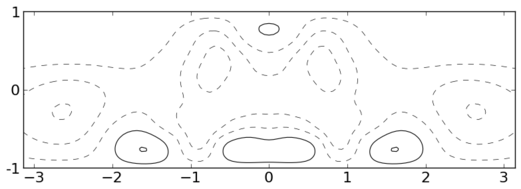} \\
\end{tabular}
\caption{{\bf Streamwise-averaged roll-streak structure of spanwise-localized traveling waves of 
channel flow.}
(a) Quiver plot of $x$-average $[v,w]$ and (b) contour plot of $x$-average $u(y,z)$
for TW2-1. (c,d) Same for TW2-2. $\alpha = 1, \Rey=400$ for both solutions. Contour lines are 
plotted at levels $u= \{-0.25, -0.15, -0.05, 0.05, 0.15\}$, with negative values in dashed 
lines and positive in solid. Quiver plots are autoscaled. The $z \in [-\upi,\upi]$ subset 
of the full $[-3\upi, 3\upi]$ computational domain is shown. 
}
\label{f:spanwise_channel_xavg}
\end{figure}

In this section we construct spanwise-localized traveling wave solutions of channel 
flow by windowing the spanwise-periodic travelings waves of \refsec{s:periodic_channel}
TW1 and TW2 in different spatial phases. The resulting solutions are 
illustrated in \reffig{f:spanwise_channel}, \reffig{f:spanwise_channel_xsect}, and 
\reffig{f:spanwise_channel_xavg}.

{\bf TW1-1}, shown in \reffig{f:spanwise_channel}(a), was formed by phase-shifting TW1 in $z$ 
by $\upi/4$ to give it symmetry group $\langle \sy, \sz \tx, \txz \rangle$, extending this 
periodic solution from a $2\upi,1\upi$ to a $2\upi,6\upi$ box, windowing the extended
periodic solution, and then applying Newton-Krylov-hookstep refinement to the windowed 
initial guess, producing a spanwise-localized traveling wave with $\langle \sy, \sz \tx, \rangle$
symmetry. The windowing parameters $a,b = 0.5,0.4$ were chosen to isolate the 
central vortex structures in swirling-strength plots. It did not take a great deal 
of effort to find windowing parameters that gave a successful initial guess; we merely 
adjusted $a$ and $b$ by tenths until swirling-strength plots of the initial guess appeared 
similar to the resultant solution in \reffig{f:spanwise_channel}(a). The initial guess for
the wave speed was set to the wave speed of the underlying periodic solution TW1, and the
pressure gradient was held fixed at $dP/dx = -2/\Rey$ where $\Rey=2000$. The Newton-Krylov-hookstep 
search converged to machine precision in six steps, consuming about half an hour of 
single-core CPU time on the machine described in \refsec{s:spanwisePCF_computation},
for a $20\!\times\!81\!\times\!256$ discretization of the $2\upi,6\upi$ computational domain.
The computed solution turned out to be smoother in $x$ and $z$ than the periodic solution
on which it was based, enough that $10^{-6}\!\times\!10^{-10}\!\times\!10^{-6}$ truncation 
levels were retained on a reduced grid of $12\!\times\!81\!\times\!192$. Similarly {\bf TW1-2} 
\reffig{f:spanwise_channel}(b) was found by refining an initial guess formed from windowing 
TW1 in in its $z$-phase with $\langle \sy, \sz, \txz \rangle$ symmetry and with windowing 
parameters $a,b = 1.2, 0.4$, resulting in a traveling wave solution with $\langle \sy, \sz \rangle$ 
symmetry.

{\bf TW2-1} and {\bf TW2-2}, shown in \reffig{f:spanwise_channel}(c,d), are likely the most 
interesting solutions presented in this paper, as they represent traveling wave solutions
that are spanwise localized and strongly concentrated near a single wall, and as such are 
the solutions most likely to be relevant to the lambda-vortex coherent structures that occur 
near the walls of shear flows (\cite{SaikiJFM93}). TW2-1 was formed by windowing 
TW2 in its $z$-phase with symmetry $\langle \tx \sz, \txz \rangle$ and windowing parameters 
$a,b = 0.6, 0.4$ to get an initial guess with $\tx \sz$ symmetry, and refining that with 
Newton-Krylov-hookstep. We were unable to form TW2-2 by the phase-shifting, windowing, and 
refining procedure employed for other localized solutions. Instead TW2-2 was formed by shifting 
TW2-1 leftwards in $z$ until its right-hand high-speed streak was centered on $z=0$, extending 
the shifted field on $z \leq 0$ onto $z \geq 0$ by $z$-mirror symmetry, and then refining
this initial guess with Newton-Krylov-hookstep. We note that TW2-1 and TW2-2 required finer 
grids to resolve the stronger vortex structure adequately compared to TW1-1 and TW1-2. Larger 
computational domains were also necessary, due to the increased width of the streamwise 
velocity  deficit (visible as contour lines in the back plane of \reffig{f:spanwise_channel}). 
The computations of TW2-1 and TW2-2 from initial guess to took about seven CPU-hours each. 

Each of the traveling wave solutions depicted in \reffig{f:spanwise_channel} is formed 
from variations of very similar basic structure, which mostly clearly seen in isolation in 
TW2-1 in \reffig{f:spanwise_channel}(c).  Near the lower wall there is an $x$-periodic 
chain of concentrated vortices, alternating in sign of circulation (clockwise/counter-clockwise), 
and nearly aligned with the $x$-axis but tilting slightly in the wall-normal and spanwise 
directions, as shown in \reffig{f:spanwise_channel_xsect}(a-e). The tilting of the chain 
of alternating vortices results in a nonuniform $x$-average in the cross-stream flow, 
specifically, a pair of counter-rotating mean vortices near the wall, 
\reffig{f:spanwise_channel_xavg}(a). The mean vortices draw low-speed fluid upwards 
between them and high-speed fluid downward on either side, producing the 
mean high-speed streaks on either side of the mean vortex pair, depicted by solid contour 
lines in \reffig{f:spanwise_channel_xavg}(b). The streamwise momentum exchanged induced by 
the near-wall mean vortices has a net negative effect: the region of mean streamwise flow 
slower than laminar indicated by negative (dashed) contour lines in \reffig{f:spanwise_channel_xavg}(b)
is larger in both $y,z$ area and magnitude than the high-speed streaks that outpace laminar
flow, indicated by solid contour lines. Thus the net effect of the roll-streak structure
is a decrease of bulk flow relative to laminar for a fixed pressure gradient. 
TW2-2 roughly consists of two copies of TW2-1's basic structure, repeated with 
mirror-symmetry about the $z=0$ plane. This is evident from comparison of TW2-1 and 
TW2-2 in \reffig{f:spanwise_channel}(c) and (d), \reffig{f:spanwise_channel_xsect}(a-e) 
and (f-g), and \reffig{f:spanwise_channel_xavg}(a,b) and (c,d). TW1-1 and TW1-2 have 
similar structure to TW2-1 and TW2-0, but with mirror symmetry in $y$ and weaker 
magnitudes of vorticity (by roughly a factor of two) and net velocity deficit relative 
to laminar (by a factor of four or more). 

The cross-stream quiver plots of these exact wall-bounded traveling waves,
\reffigs{f:spanwise_channel}{f:spanwise_channel_xavg} bear intriguing resemblances 
to the cross-sections of near-wall lambda vortices of developing channel flow
shown in \cite{SaikiJFM93} figures 10 and 11. For example, the streamwise mean $v,w$ 
flow of TW2-1 in \reffig{f:spanwise_channel_xavg} is strikingly like the temporal-mean 
$v,w$ of the K-type-generated lambda vortex shown in  \cite{SaikiJFM93} figure 10(b). 
The $\Lambda$-structure shown in \cite{SaikiJFM93} figure 2(b) is evident in 
TW2-2 in \reffig{f:spanwise_channel}(d). Of course the traveling waves in 
this paper are exact solutions of different flow conditions than the 
spatially-developing channel flow of \cite{SaikiJFM93}, so some translation 
of the flow conditions is required to make more precise comparisons. We intend to 
pursue this in future research.

\begin{table}
\begin{tabular}{lllllllll}
      & symmetry & $\alpha, \gamma$ & $\Rey$ & $c$ & $L_x,L_z$ & ~~~grid & accur. & tails \\
TW1  & $\langle \sy, \sz, \txz \rangle$ & $1,2$ & 2300 & 0.673 & $2\upi,\upi$  & $16\!\times\!81\!\times\!32$   & $10^{-7}$ &  -\\
TW1-1 & $\langle \sy, \tx \sz \rangle$   & $1,-$ & 2300 & 0.674 & $2\upi,6\upi$ & $12\!\times\!81\!\times\!192$ & $10^{-6}$ & $10^{-5}$\\
TW1-2 & $\langle \sy, \sz \rangle$       & $1,-$ & 2300 & 0.674 & $2\upi,6\upi$ & $12\!\times\!81\!\times\!192$ & $10^{-6}$ & $10^{-5}$\\
TW2   & $\langle \sz, \txz \rangle$      & $1,2$ & 2300 & 0.564 & $2\upi,\upi$  & $24\!\times\!97\!\times\!48$   & $10^{-6}$ & - \\
TW2-1 & $~\tx \sz $                       & $1,-$ & 2300 & 0.661 & $2\upi,6\upi$ & $16\!\times\!97\!\times\!216$ & $10^{-6}$ & $10^{-5}$\\
TW2-2 & $~\sz$                            & $1,-$ & 2300 & 0.648 & $2\upi,6\upi$ & $16\!\times\!97\!\times\!216$ & $10^{-6}$ & $10^{-5}$\\
TW3  & $\langle \sz, \txz \rangle$      & $1,6$ & 4000 & 0.475 & $2\upi,\upi/6$ & $24\!\times\!109\!\times\!24$ & $10^{-6}$ & -\\
\end{tabular}
\caption{\bf Characteristics of traveling-wave solutions of channel flow.}
\label{t:channeltw_char}
\end{table}

\refTab{t:channeltw_char} summarizes physical characteristics and discretization 
properties of the computed traveling-wave solutions of channel flow. Wave speed
$c$ is in nondimensionalized units where the centerline velocity of laminar 
flow is $\bar{U} = 1$. $L_x, L_z$ are the size of the computational domain. 
Accuracy is measured as $\| \bff^T(\bu)) - \tau \bu \|/ T$ for $T=10$, 
where $\bff^T$ is the time-$T$ forward time integration of the Navier-Stokes as
performed by direct numerical simulation, and where $\tau = \tau(cT,0)$ is the 
$x$-translation resulting from the computed wave speed $c$ over time $T$. The magnitude of 
the tails is given as the maximum of $|u|, |v|$ and $|w|$ over $x,y$ at $z=\pm L_z/2$,
the edge of the spanwise-extended periodic box.

\section{Discussion}
\label{s:discussion}

\subsection{Exponential decay of tails}

The spanwise localized solutions presented in this paper display a 
three-part structure: a core region that closely resembles a periodic 
solution, a transition region, and weak tails that decay to laminar 
flow. In this section we show that the tails of spanwise-localized 
streamwise-periodic equilibria
are dominated by a mode that decays exponentially at $e^{-\alpha |z|}$, 
where $\alpha$ is the fundamental streamwise wavenumber, 
and that the structure of the tails depends on flow parameters $\alpha, 
\Rey$, the laminar flow profile $U(y)$, and the wavespeed $c$, but not on the details of the 
solution's core region. 

As the tails of a spanwise-localized equilibrium or traveling wave approach laminar flow, 
the perturbation velocity $\bu$ approaches zero, so we expect $\bu$ 
to approximately satisfy the linearized form of \refeq{eq:NSE_PCF}
in which $\bu \cdot \grad \bu$ is set to zero,
\begin{equation} 
\frac{\partial \bu}{\partial t} + U \frac{\partial \bu}{\partial x} + v \: U'\: \be_x  
  = -\nabla p  + \frac {1} {\Reynolds} \nabla^2 \bu, \quad \nabla \cdot \bu  = 0.
\label{eq:NSE_linearized}
\end{equation}
We look for normal-mode solutions of the form
\begin{align}
 \bu_{j,\gamma}(\bx) =  \tbu_{j,{\gamma}}(y) \: e^{i(j \alpha (x-ct) + \gamma z)}, \quad
   p_{j,\gamma}(\bx) =   \tp_{j,{\gamma}}(y) \: e^{i(j \alpha (x-ct) + \gamma z)}  
\label{eq:tails_ansatz}
\end{align}
where $\alpha$ is real and $\gamma = \gamma_r + i\gamma_i$ has $\gamma_i > 0$ for tails 
that decay exponentially as $z \rightarrow \infty$. The slowest decaying normal mode solution, with the 
smallest positive $\gamma_i$, will dominate the tails as $z \rightarrow \infty$. For the 
remainder of this section we drop the $j,\gamma$ subscripts. To eliminate pressure 
we convert to velocity-vorticity form by taking the $y$-components of the curl 
and the curl of the curl of \refeq{eq:NSE_linearized} 
\begin{align}
\left( \frac{\partial}{\partial t} + U\frac{\partial}{\partial x}\right) \eta + U' v_z &= \frac{1}{\Rey}\nabla^2 \eta, \nonumber \\ 
\left( \frac{\partial}{\partial t} + U\frac{\partial}{\partial x}\right) \nabla^2 v - U'' v_x &= \frac{1}{\Rey}\nabla^4v,
\label{eq:velocity_vorticity}
\end{align}
where $\eta = u_z - w_x$ is the wall-normal 
vorticity. Boundary conditions are $\eta(x,\pm 1,z) = 0$ and 
$v(x,\pm 1,z) = v'(x,\pm 1,z) = 0$.

\Eqn{eq:velocity_vorticity} and boundary conditions permit a number of types
of solution. Where symmetries allow, the solution that dominates behavior in the tails of 
localized plane Couette equilibria and traveling waves turns out to be the trivial solution 
$v = \eta = 0$. This solution in conjunction with the divergence-free condition 
and the ansatz \refeq{eq:tails_ansatz} requires that $\tu_{xx} = -\tu_{zz}$ and 
$\tw_{xx} = -\tw_{zz}$, which is satisfied by $\gamma = \pm i j\alpha$ and 
$\tw = \pm i j \tu$, the different signs governing exponential decay in the 
different limits $z \rightarrow \pm \infty$. The $j=1$ mode with 
$\gamma =  \pm i \alpha$ and $\tw = \pm i \tu$ thus gives the
slowest exponential decay rate ($j=0$ is ruled out since it does not decay and 
thus cannot be part of a $z$-localized solution). The $y$ component of \refeq{eq:NSE_linearized} 
gives that $\partial \tp / \partial y = 0$ so that $\tp(y) = \tp$ is a complex 
constant whose magnitude and phase are set by the pressure conditions at the 
edge of the transition region at some fixed value of $z$. The $x$ component of 
\refeq{eq:NSE_linearized} gives
\begin{align}
\tu'' - i \alpha \Rey(U - c) \tu = i \alpha \Rey \; \tp.
\label{eq:utails_ODE}
\end{align}
The boundary conditions $\tu(\pm 1) = 0$ set the homogeneous solution to zero,
so that $\tu$ and thus $\tw$ are determined by the fixed value of $\tp$, the base flow profile $U(y)$, 
and the parameters $\alpha$, $c$ and $\Rey$. They are independent of the structure of
the core-region solution (except for differences of $x$ phase between the $\pm z$ tails
resulting from the symmetries of the solution). Thus the $v = \eta = 0$ solution 
to the linearized Navier-Stokes equations contributes to the tails an
exponentially decaying mode of form $[\tu(y), \: 0, \: i\tu(y)]  e^{i \alpha (x-ct) - \alpha z}$
as $z \rightarrow \infty$.

\begin{figure}
\begin{center}
\begin{tabular}{cc}
{\footnotesize (a)} \hspace{-1.8mm} \includegraphics[width=0.45 \textwidth]{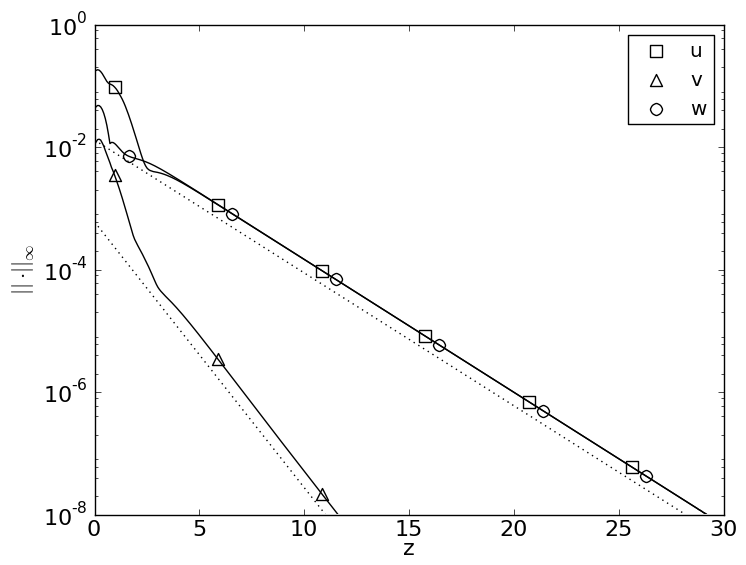} &
{\footnotesize (b)} \hspace{-1.8mm} \includegraphics[width=0.45 \textwidth]{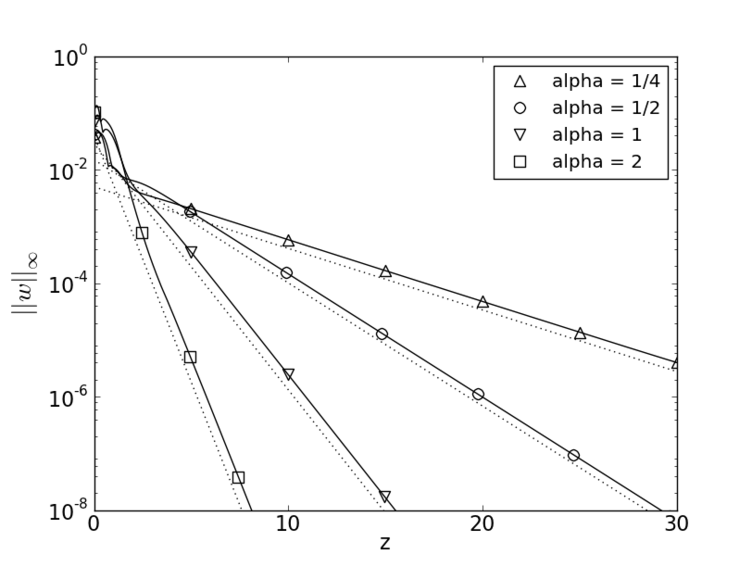} 
\end{tabular}
\end{center}
\caption{{\bf Exponential decay in the tails of spanwise localized plane Couette equilibria.}
(a) Componentwise decay rates for EQ7-1 at $\alpha = 1/2$, $\Rey = 600$: $u,v,w$ (solid lines) 
scale as $e^{-\alpha z}, e^{-2\alpha z}, e^{-\alpha z}$ (dotted lines).
(b) Decay of $\|w\|_{\infty}$ for EQ7-1 (solid lines) compared to $e^{-\alpha z}$ (dotted lines)
at $\Rey=600$ and several values of $\alpha$.
}
\label{f:EQ7tails}
\end{figure}

\begin{figure}
\begin{tabular}{cc}
\begin{minipage}{.45\textwidth}
{\footnotesize (a)} \hspace{-1.7mm} \includegraphics[width= \textwidth]{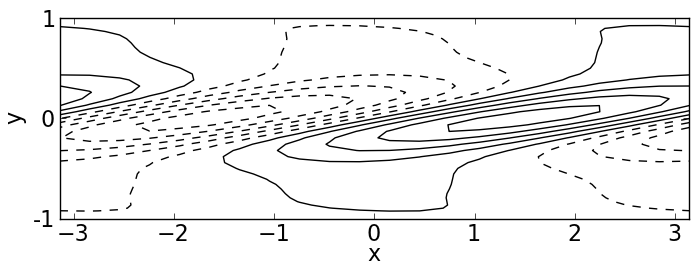}
{\footnotesize (b)} \hspace{-1.7mm} \includegraphics[width= \textwidth]{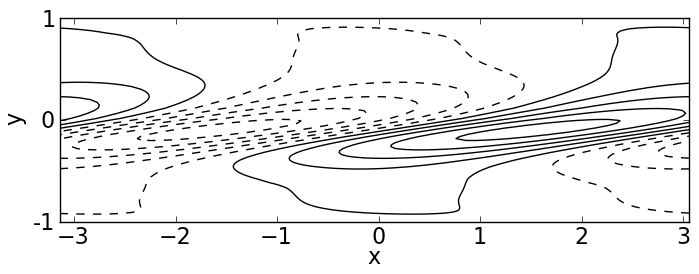}
\end{minipage} &
\begin{minipage}{.45\textwidth}
{\footnotesize (c)} \hspace{-1.7mm} \vspace{-2mm} \includegraphics[width= \textwidth]{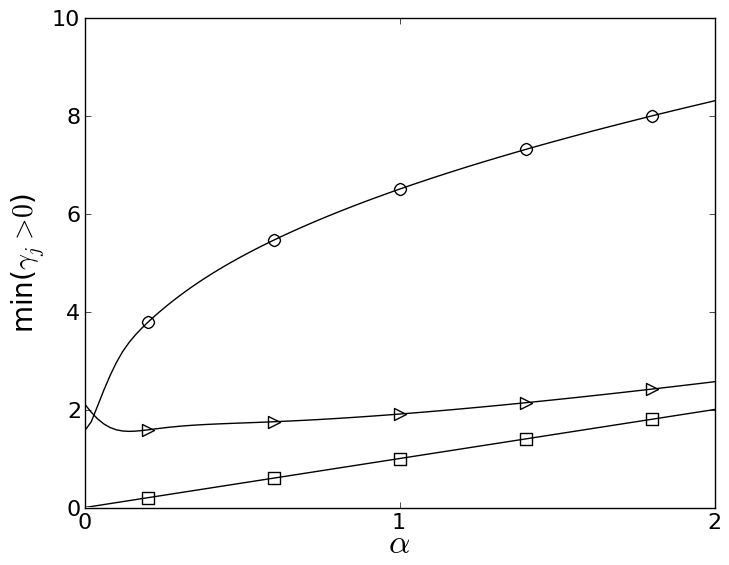}
\end{minipage} \\
\end{tabular}
\caption{{\bf (a,b) Asymptotic form of the tails of spanwise-localized equilibria of plane Couette flow.} 
Contours of streamwise velocity $u(x,y)$ at fixed $z$: (a) a slice of EQ7-1 for $\alpha = 1$ and 
$\Rey = 400$, at $z=11$, where $||u||_{\infty} \approx 10^{-6}$, and (b) the asymptotic $v=\eta=0$ 
normal mode solution. 
Contours are plotted at $\pm [0.15, 0.45, 0.75]$ times the maximum of $u$, with negative values in
dashed lines. 
{\bf (c) Exponential decay rate for three types of normal modes in the tails} of
plane Couette equilibria at $\Rey = 400$. The minimal $\gamma_i$ as a function of $\alpha$ is 
plotted for ($\square$) $v=\eta=0$ solutions, ($\circ$) solutions of (\ref{eq:etaEig}) with $\tv(y)=0$, 
and ($\triangleright$) solutions of (\ref{eq:vEig}). 
}
\label{f:xyslices}
\end{figure}

\refFig{f:EQ7tails} confirms that over a wide range of $\alpha$, the tails of EQ7-1 
are dominated by $u,w$ components that decay as $e^{-\alpha z}$. \refFig{f:EQ7tails}(a) 
shows the inf-norm of $u,v$ and $w$ as a function of $z$ for EQ7-1 at $\alpha = 1$.
In this context $\|u\|_{\infty}(z)$ is the maximum of $|u(x,y,z)|$ over $x,y$ as a 
function of $z$. As argued above, the magnitudes of the $u$ and $w$ components are 
equal in the tails and scale as $e^{-\alpha z}$. The higher-order $e^{-2 \alpha z}$ 
scaling for $v$ results from the quadratic nonlinear term in $u_{1,\gamma},w_{1,\gamma}$ 
that has been suppressed on the right-hand side of \refeq{eq:NSE_linearized} for the 
$v_{0,\gamma}$ equation in 
for $j=0$. 
For \reffig{f:EQ7tails}(b), we continued EQ7-1 parametrically in $\alpha$ and observed 
that the  $e^{-\alpha z}$ scaling holds over the explored range of $1/4 \leq \alpha \leq 2$. 
For further confirmation of the dominance of $v=\eta=0$ modes, \reffig{f:xyslices}(a) 
compares an $x,y$ slice of streamwise velocity $u(x,y,z)$ from EQ7-1 at a fixed $z$ to 
(b) the same as computed from \refeq{eq:utails_ODE}. 

The linearized equations for the tails have solutions other than $v = \eta = 0$; to 
show these decay more rapidly than the $v = \eta = 0$ modes (for plane Couette flow), we reduce \eqn{eq:velocity_vorticity}
to the eigenvalue equations
\begin{align}
i\alpha (U-c)\teta + i\gamma U' \tv  &= \frac{1}{\Rey}(\teta'' - (\alpha^2+\gamma^2)\teta), \label{eq:etaEig}\\
i\alpha (U-c)(\tv'' - (\alpha^2 + \gamma^2)\tv) - i\alpha U''\tv &= \frac{1}{\Rey}(\tv'''' - 2(\alpha^2 + \gamma^2)\tv'' + (\alpha^2 + \gamma^2)^2\tv). \label{eq:vEig}
\end{align}
The latter is the time-independent form of the familiar Orr-Sommerfeld equation for 
three-dimensional disturbances. Equation (\ref{eq:vEig}) is independent of $\teta$ 
and the eigenvalues $\gamma$ can be found numerically for given $\alpha$, $c$, $\Rey$,
and $U(y)$. The $i\gamma U'\tv$ coupling 
term in (\ref{eq:etaEig}) acts as a nonhomogeneous forcing, requiring particular 
solutions for $\teta(y)$ to match the eigenmodes of (\ref{eq:vEig}). Eigenvalues 
distinct from those found for (\ref{eq:vEig}) can be found by solving (\ref{eq:etaEig}) 
with $\tv(y) = 0$. \refFig{f:xyslices}(c) shows the minimal $\gamma_i$ allowed 
by these two equations as a function of $\alpha$ for plane Couette equilibria at
$Re=400$. Note that $v=\eta=0$ modes have the smallest $\gamma_i$ for $\alpha$ in the 
range shown ($\alpha < 2$); thus these modes dominate the behavior of the tails in all 
solutions with streamwise wavelength greater than $\upi$. It should be noted, however, 
that streamwise constant ($\alpha = 0$) modes can exist in domains of any length $L_x$, 
and thus might play a dominant role in short (large $\alpha$) domains if the symmetries
permit them.

\subsection{Asymptotic scaling of streamwise Fourier harmonics}
\label{s:scaling}

In this section we provide a numerical account of the large-Reynolds behavior
of the periodic and localized solutions developed in previous sections. In 
particular we measure the scaling of various streamwise Fourier components of 
the solutions with $\Rey$, and we show the development of critical layers at 
large $\Rey$. These features are key to the asymptotic analysis of NBCW suggested 
by \cite{WangPRL07} and developed into a complete theory by \cite{HallJFM10}. 
The main results are as follows. The streamwise Fourier components of EQ7 and 
solutions related to it, localized and in channel conditions, obey scaling 
laws similar to those of NBCW, albeit with different exponents and substantially 
different magnitudes, suggesting that \cite{HallJFM10}'s asymptotic analysis
could be carried over to the new solutions. EQ8 and the $y$-asymmetric channel 
flow solutions, in contrast, do not fit the asymptotic scaling framework so 
cleanly. All solutions appear to have well-defined critical layers, however, and 
the critical layer is particularly simple for EQ7 and its localized counterparts.

As suggested by \cite{WangPRL07} and developed into a complete theory by 
\cite{HallJFM10}, a reduced PDE system can be developed for the spatially 
periodic NBCW solution from an asymptotic analysis of its streamwise Fourier 
modes and the critical layer that develops at large Reynolds numbers. 
\cite{WangPRL07} showed numerically that NBCW has a simplified, quasi-2D 
structure in the limit of large Reynolds numbers, with a balance between 
an $O(1)$ streamwise-constant streaks, $O(\Rey^{-1})$ streamwise-constant rolls, and 
an $O(\Rey^{-0.9})$ mode in the first (fundamental) streamwise Fourier harmonic, which 
concentrates in a critical layer of thickness $O(\Rey^{-1/3})$. \cite{HallJFM10} in 
turn developed an asymptotic theory for NBCW based on vortex-wave interaction that 
provides insight to the physics of how these components of NBCW balance, predicts 
their scaling exponents, and which reduces the computation of the solution from 
a 3D Navier-Stokes problem at large-$\Rey$ to a 2D PDE at $\Rey=1$ coupled with a 
linear wave evolution equation. Specifically, the interactions of very small 
fundamental-mode streamwise waves within the critical layer generate nonzero mean 
stresses that cause jumps in the pressure and the normal derivative of roll velocity 
across the critical layer. The jump in roll shear drives the mean rolls, which in 
turn drive the mean streaks. \cite{HallJFM10} show that these effects balance
to leading order in $\Rey^{-1}$, and that the asymptotic theory also reduces computation 
of the 3D steady state at high Reynolds number to a simpler 2D calculation at unit 
Reynolds number coupled with a linear wave evolution equation. 

This reduced quasi-2D PDE model of \cite{HallJFM10} is of particular interest 
to us since a theoretical analysis of spanwise localization in solutions of 
Navier-Stokes should be easier to develop in the context of a reduced model. 
There is strong numerical evidence that a theory of localization in solutions 
of Navier-Stokes might be developed. \cite{SchneiderJFM09} noted a remarkable 
resemblance between the $x,y$-averaged energy of the localized NBCW solutions 
and localized solutions of the 1d Swift-Hohenberg equation found by \cite{BurkeChaos07}. 
The similarity was made more remarkable by \cite{SchneiderPRL10}'s demonstration 
that the localized NBCW solutions undergo homoclinic snaking under continuation in 
Reynolds number, just as the localized Swift-Hohenberg solutions do under 
continuations in their bifurcation parameter. For Swift-Hohenberg, homoclinic 
snaking of localized solutions is quite well understood theoretically via
``spatial dynamics'' (\cite{BurkeChaos07}). Time independence reduces the 
4th-order Swift-Hohenberg PDE for $u(x,t)$ to 4th-order ordinary differential 
equation (ODE) on $u(x)$, which can then considered as a 4-dimensional
dynamical system where $x$ plays the role of time. In this view, spatially 
periodic solutions correspond to periodic orbits of the spatial dynamics, and 
spatially localized solutions correspond to homoclinic orbits that start at the 
origin $u=0$ at $t\rightarrow -\infty$, grow away from it along unstable direction, 
wander at finite amplitude for some time, and then reapproach $u=0$ at 
$t\rightarrow \infty$ along a stable direction. Localized solutions display 
approximately periodic form in their core regions when the finite-amplitude 
excursion away from $u=0$ makes a number of circuits in the neighborhood of an 
unstable periodic orbit of the spatial dynamics. 
Thus the close correspendence between localized Navier-Stokes solutions and 
localized Swift-Hohenberg solutions points to the possibility of a theoretical
explanation of localization in invariant solutions of Navier-Stokes. It also 
points to the importance of a reduced PDE system describing the localized 
solutions, ideally to a 1d system in the spanwise coordinate, to which the 
idea of spatial dynamics might be applied. We report on the scaling of streamwise 
Fourier components and the  development of a critical layer in the plane Couette 
and channel solution because they are essential ingredients for developing such 
a reduced-order PDE model. 

The perturbation velocity field of equilibrium and traveling wave solutions 
can be expressed as a sum of streamwise ($x$) Fourier modes in the form
\begin{align}
\bu(\bx,t) = \sum_{j} \hbu_j(y,z) e^{i j \theta}  
\end{align}
where $\theta = \alpha(x-ct)$ and where $\hbu_{-j}(y,z) = \hbu_{j}^*(y,z)$ so that
$\bu$ is real-valued. The streamwise constant mode $\hbu_0$ can be decomposed into 
streamwise {\em streaks} $\hu_0(y,z)$ and cross-stream {\em rolls} $[0, \hv_0, \hw_0] (y,z)$. 
Recall that $\bu$ is the deviation from laminar flow $U(y) \be_x$, with 
$\butot = \bu + U(y) \be_x$ and $\hutotO = \hu_0 + U(y)$, so that the streaks are 
defined relative to laminar flow. \cite{WangPRL07} define streaks relative to the 
$z$-averaged mean flow, but we do not, since $z$-averaging is inappropriate for 
spanwise-localized solutions. \cite{HallJFM10} refer to the first (fundamental) 
harmonic $\hbu_1(y,z)$ or $\hbu_1(y,z) \exp(i \alpha (x-ct)) + \cc$ as the {\em wave} 
mode. 

\begin{figure}
\begin{tabular}{ccc}
{\footnotesize (a)} \hspace{-2mm} \includegraphics[width=0.3\textwidth]{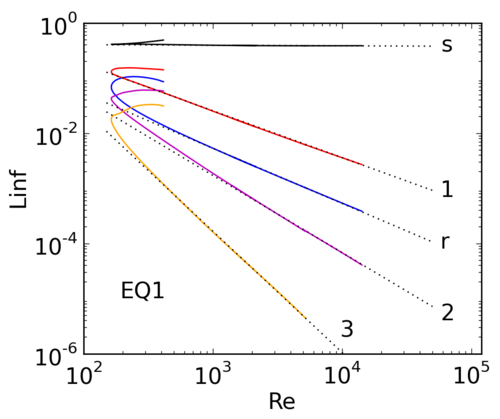} & \hspace{-2mm}
{\footnotesize (b)} \hspace{-2mm} \includegraphics[width=0.3\textwidth]{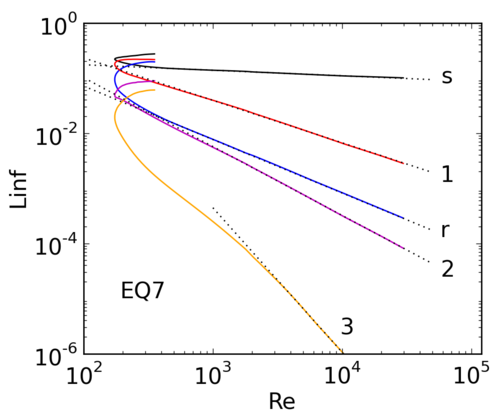}   \hspace{-2mm}
{\footnotesize (c)} \hspace{-2mm} \includegraphics[width=0.3\textwidth]{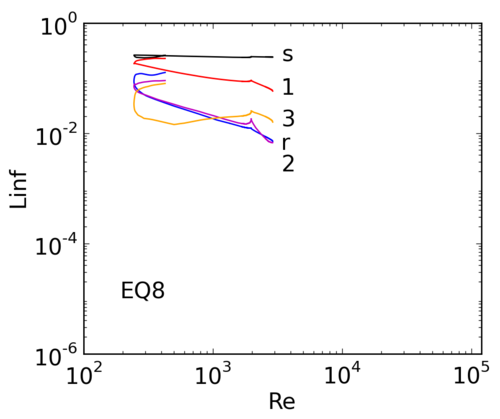} \\
{\footnotesize (d)} \hspace{-2mm} \includegraphics[width=0.3\textwidth]{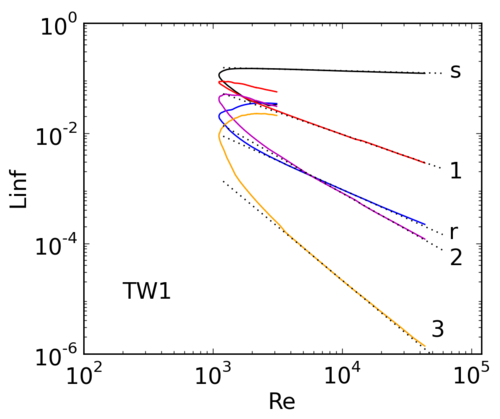} & \hspace{-2mm}
{\footnotesize (e)} \hspace{-2mm} \includegraphics[width=0.3\textwidth]{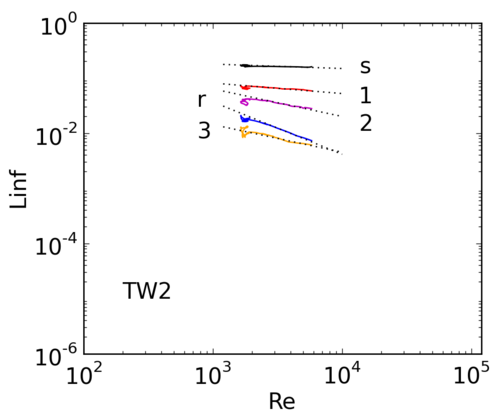}   \hspace{-2mm}
{\footnotesize (f)} \hspace{-2mm} \includegraphics[width=0.3\textwidth]{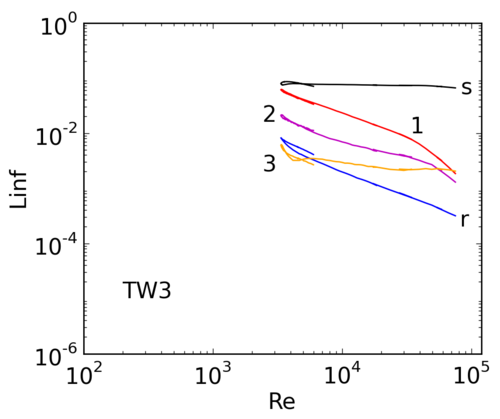} \\
{\footnotesize (g)} \hspace{-2mm} \includegraphics[width=0.3\textwidth]{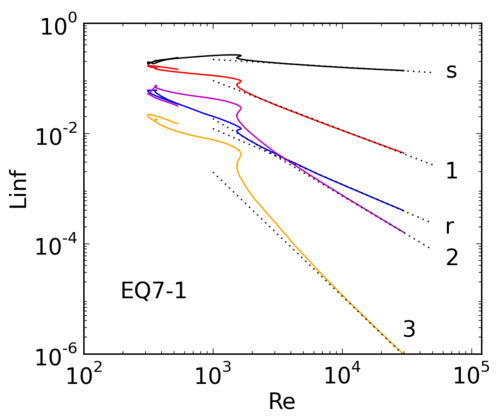} & \hspace{-2mm}
{\footnotesize (h)} \hspace{-2mm} \includegraphics[width=0.3\textwidth]{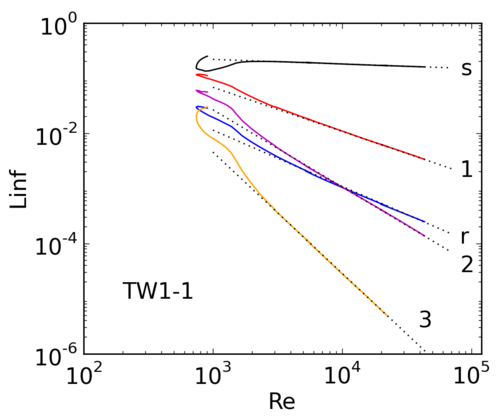}          \hspace{-2mm}
{\footnotesize (i)} \hspace{-2mm} \includegraphics[width=0.3\textwidth]{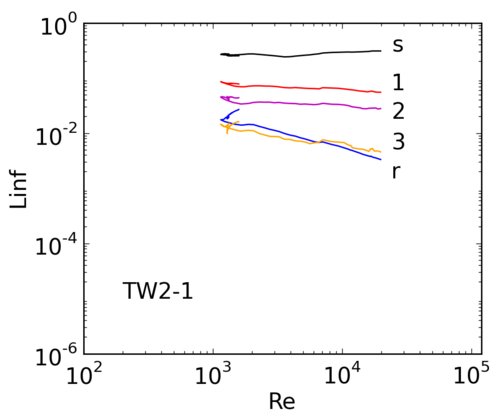} \\
\end{tabular}
\caption{{\bf Scaling of streamwise Fourier modes of plane Couette and channel solutions.}
(a,b,c) Spatially periodic plane Couette equilibria NBCW, EQ7, EQ8, at $\alpha,\gamma = 1,2$.
(d,e,f) Spatially periodic channel flow traveling waves TW1, TW2, TW3, at $\alpha,\gamma = 1,2$.
(g,h,i) Spanwise localized solutions: EQ7-1 plane Couette equilibrium and
TW1-1 and TW2-1 channel flow traveling wave, at $\alpha = 1$. The magnitude of 
various Fourier components of the velocity, as measured by inf-norm is plotted 
against Reynolds number. The labels s,r,1,2,3 indicate the streaks $\hu_0$, rolls 
$[\hv_0,\hw_0]$, and the 1st, 2nd, and 3rd streamwise Fourier harmonics 
$\hbu_{1}$, $\hbu_{2}$, and $\hbu_{3}$. The vertical axis is the 
$\infty$-norm of the given component (see text).
}
\label{f:scaling}
\end{figure}

\newcolumntype{.}{D{.}{.}{3}}
\begin{table}
\centering
\begin{tabular}{ll|.......}
$j$   &          & \multicolumn{1}{c}{NBCW} &  
\multicolumn{1}{c}{EQ7} &  \multicolumn{1}{c}{EQ7-1} & \multicolumn{1}{c}{EQ7-2} &
\multicolumn{1}{c}{TW1} &  \multicolumn{1}{c}{TW1-1} & \multicolumn{1}{c}{TW1-2} \\
\hline 
$0$   & streak   & 0.01   & 0.10  & 0.14  & 0.12  & 0.06 & 0.08  & 0.09 \\
$1$   & wave     & 0.85   & 0.77  & 0.90  & 0.83  & 0.80 & 0.80  & 0.82 \\
$0$   & roll     & 1.0    & 0.97  & 1.0   & 1.0   & 1.05 & 1.02  & 1.00 \\
$2$   &          & 1.4    & 1.25  & 1.4   & 1.35  & 1.35 & 1.4   & 1.35\\
$3$   &          & 2.2    & 2.6   & 2.25  & 2.45  & 1.95 & 2.2   & 2.0
\end{tabular}
\caption{{\bf Scaling exponents for streamwise Fourier harmonics of equilibrium 
and traveling wave solutions of plane Couette and channel flow.} Components
of solutions $\bu$ scale as $|| \cdot ||_{\infty} = O(\Rey^{-\mu})$ for the given 
values of $\mu$.
}
\label{t:exponents}
\end{table}

\refFig{f:scaling} shows the scaling with Reynolds number of the magnitude of the 
streaks, rolls, waves, and second and third harmonics for a number of plane Couette 
and channel solutions. The magnitude is computed with the inf-norm; for example, 
the magnitude $|| \hbu_1||_{\infty}$ of the wave is the maximum over $x,y,z$ 
and the vector components $[u,v,w]$ of $\bu = \hbu_1(y,z) e^{i \alpha x} + \hbu_{-1}(y,z) 
e^{-i \alpha x}$. A number of these solutions exhibit very clear scaling of the form 
$|| \hbu_j ||_{\infty} = O(\Rey^{-\mu_j})$. For example \refFig{f:scaling}(a) shows
that for NBCW, the streaks are $O(1)$, the rolls $O(\Rey^{-1})$, and the waves 
$O(\Rey^{-0.85})$. The streak and roll values scalings equal the theoretical predictions
of \cite{HallJFM10}, and the wave scaling is within 2\%. Note that the choice of 
inf-norm changes the scaling exponent for the NBCW wave component compared to the value 
of $\mu_1 = 0.9$ reported by \cite{WangPRL07}. 

EQ7 in \reffig{f:scaling}(b) shows equally clear asymptotic scaling with exponents 
comparable but not equal to those of NBCW. The same is true of all solutions derived
from EQ7 by parametric continuation and localization by windowing. Examples of such
EQ7-related solutions are shown in \reffig{f:scaling}, namely (d) TW1, the spatially 
periodic traveling wave of channel flow obtained from EQ7 by continuation; (g) EQ7-1,
the spanwise localized equilibrium of plane Couette flow obtained by windowing; and
(i) TW-1, the spanwise localized traveling wave of channel flow obtained by windowing
TW-1. Scaling exponents for these solutions plus EQ7-2 and TW7-2 are listed in 
\reftab{t:exponents}. It should be noted that the magnitudes of Fourier harmonics of
EQ7 and solutions derived from it are substantially different from those of NBCW. 
For example, comparison of EQ7 in \reffig{f:scaling}(b) to NBCW in (a) shows that 
EQ7's streaks are about a factor of three smaller than NBCW's, and its fundamental
harmonic is about a factor of three larger, resulting in an order of magnitude less
scale separation between these components. 

In contrast, the upper-branch and $y$-asymmetric solutions reported here have no clear 
asymptotic scaling in streamwise Fourier harmonics and substantially poorer separation 
of scales, namely \reffig{f:scaling}(b) EQ8, the upper branch of EQ7; (e) TW2, a spanwise 
periodic traveling wave of channel flow obtained from judiciously chosen DNS data;
(f) TW3, a higher-wavenumber spanwise periodic traveling wave of channel flow obtained 
from continuation in $\Rey$ of TW2; and (i) TW2-1, a spanwise localized traveling wave of 
channel flow obtained from windowing TW2. TW2-2 (not shown) is similar to TW-1. Among 
these, none of EQ8, TW2, TW2-1, or TW-2 continue in a straightforward fashion to higher 
Reynolds number; instead the solutions curves turn around at finite $\Rey$ and follow 
complex paths. The same is true for EQ2, the upper branch of NBCW. The numerical evidence 
thus weighs against the possibility of an asymptotic analysis of these solutions based on
streamwise Fourier harmonics. 

\subsection{Critical layers}
\label{s:critical}

The development of critical layers is an important consequence of the 
separation of scales in the streamwise Fourier modes (\cite{WangPRL07} and 
\cite{HallJFM10}). The critical layer is the surface on which the mean streamwise 
fluid velocity matches the wavespeed, i.e. $\hutotO(y,z) = c$. When higher harmonics
become negligible and the roll velocities $\hv_0,\hw_0$ are small compared to 
the streaky streamwise velocity $\hutotO$, the equation for the fundamental mode simplifies to 
\begin{align}
\left[ i \alpha (\hutotO - c) \hbu_1 + (\hbu_1 \cdot \grad \hbutotO) \be_x \right] e^{i\theta}
= \grad (\hp_1 e^{i\theta}) + \Rey^{-1} \lapl (\hbu_1 e^{i\theta})
\label{e:critical}
\end{align}
As argued by \cite{WangPRL07}, for large $\Rey$ the fundamental harmonic $\hbu_1$ 
concentrates in a region of thickness $\delta = \Rey^{-1/3}$ about the critical layer, 
in which \refeq{e:critical} is dominated by a balance between its first and last terms. 
For a point $\bx$ in this region and $\bx_c$ nearby on the critical layer, this requires 
a balance of $\alpha (\hutotO(\bx) - c) \approx \alpha (\bx - \bx_c) \cdot \grad \hbutotO$ 
against $ \Rey^{-1} \lapl$. If $\delta \sim | \bx - \bx_c |$ is the thickness of the 
region, the balance requires $\alpha \delta |\grad \hbutotO | \sim \Rey^{-1} \delta^{-2}$
or $\delta = (\alpha |\grad \hbutotO| \Rey)^{-1/3}$.

\begin{figure}
\begin{tabular}{ccc}
 \hspace{-2mm} {\footnotesize (a)} \hspace{-2mm} \includegraphics[width=0.29\textwidth]{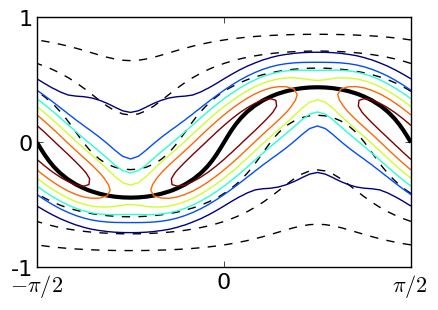} &
 \hspace{-2mm} {\footnotesize (c)} \hspace{-2mm} \includegraphics[width=0.29\textwidth]{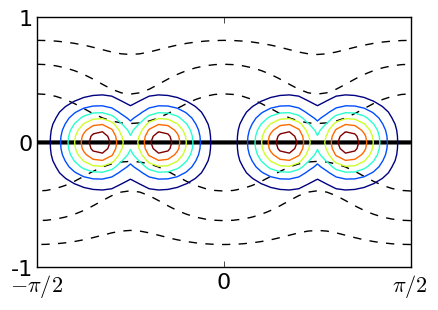} &
 \hspace{-2mm} {\footnotesize (e)} \hspace{-2mm} \includegraphics[width=0.29\textwidth]{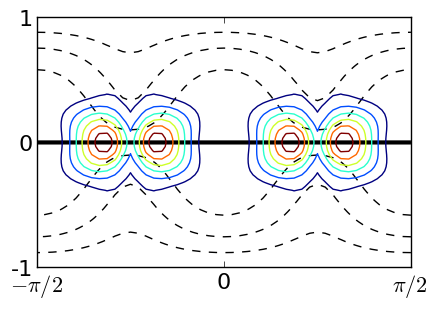} \\
 \hspace{-2mm} {\footnotesize (b)} \hspace{-2mm} \includegraphics[width=0.29\textwidth]{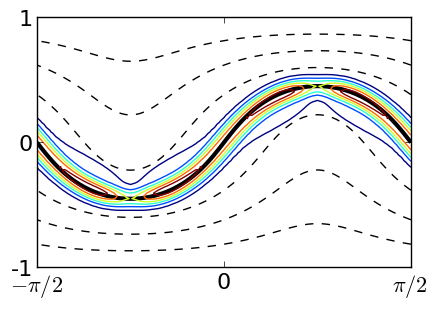} &
 \hspace{-2mm} {\footnotesize (d)} \hspace{-2mm} \includegraphics[width=0.29\textwidth]{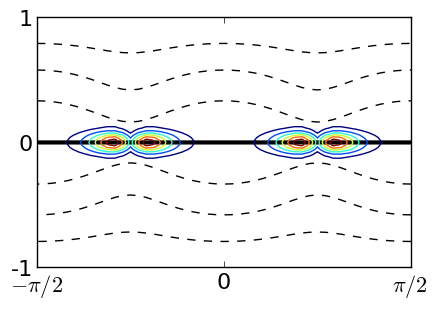} &
 \hspace{-2mm} {\footnotesize (f)} \hspace{-2mm} \includegraphics[width=0.29\textwidth]{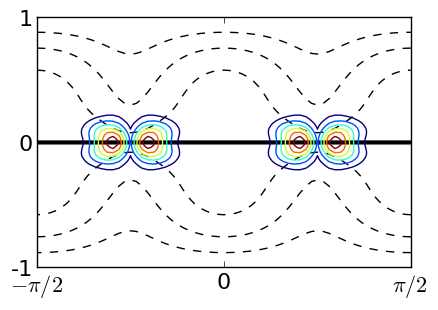} 
\end{tabular}
\caption{{\bf Critical layers of spanwise periodic equilibria of plane Couette flow.} 
(a,b) NBCW at $\Rey=1000$ and $30000$, 
(c,d) EQ7 at $\Rey=1000$ and $30000$, and
(e,f) EQ8 at $\Rey=1000$ and $3000$ 
for streamwise, spanwise wavenumbers $\alpha, \gamma = 1,2$. Dashed contour lines 
show total streamwise velocity at levels $\utot = \pm \{0, 0.25, 0.50, 0.75\}$. The 
critical layer where $\utot(y,z) = c = 0$ is shown with a thick solid contour line. 
Thin solid contour lines show the autoscaled magnitude of the fundamental Fourier 
harmonic, $|\hbu_1|$. The horizontal and vertical coordinates are $z,y$ respectively.}
\label{f:periodic_pcf_critical}
\end{figure}

\refFig{f:periodic_pcf_critical} illustrates the development of the critical layer 
for three spatially periodic equilibria of plane Couette flow. For equilibria, the 
wavespeed vanishes, so the critical layer in these plots is the surface $y=f(z)$ on
which $\utot(y,z) = 0$. For NBCW, shown in \reffig{f:periodic_pcf_critical}(a,b),
the height of the critical layer varies in $z$, and its thickness $\delta$ decreases
as $\Rey^{-1/3}$ between (a) $\Rey=1000$ and (b) $\Rey=30000$. \refFig{f:periodic_pcf_critical}(a,b)
largely duplicates figures 2 and 3 of \cite{WangPRL07}; however we note that our plots 
show contourlines of $|\hbu_1|$ rather than just the vertical component, $|\hv_1|$,
and so more clearly convey the fact that, for NBCW, the concentration of the fundamental 
mode is spread almost uniformly over the entire critical layer. 

The critical layers for EQ7 and its upper branch EQ8 are markedly different 
(\reffig{f:periodic_pcf_critical}(c,d) and (e,f)). First, the critical layer for 
these solutions is the line $y=0$. This follows from their $\sxy$ symmetry, i.e. 
$[u,v,w](x,y,z) = [-u,-v,w](-x,-y,z)$. Under this symmetry the $x$-average of 
the perturbation velocity and total velocity vanishes on $y=0$. We note that some 
of the complexity of \cite{HallJFM10}'s analysis results from the need to work 
in a coordinate system aligned with the curved criticial layer; for EQ7 and EQ8 this 
complexity would be eliminated. Second, the fundamental mode $\hbu_1$ concentrates 
not uniformly over the whole critical layer, but apparently on isolated spots within 
it. Third, EQ8 seems to form a critical layer at $\Rey=2000$, even though its scale 
separation is much poorer and its large-$\Rey$ limit apparently does not exist.

\begin{figure}
\begin{center}
\begin{tabular}{cc}
{\footnotesize (a)} \hspace{-2mm} \includegraphics[width=0.45\textwidth]{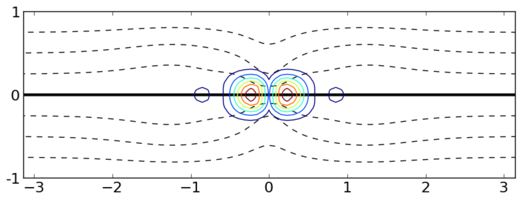} &
{\footnotesize (c)} \hspace{-2mm} \includegraphics[width=0.45\textwidth]{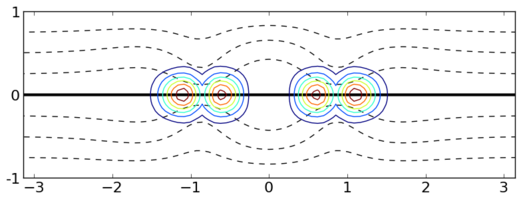} \\
{\footnotesize (b)} \hspace{-2mm} \includegraphics[width=0.45\textwidth]{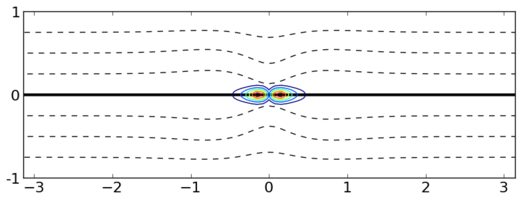} &
{\footnotesize (d)} \hspace{-2mm} \includegraphics[width=0.45\textwidth]{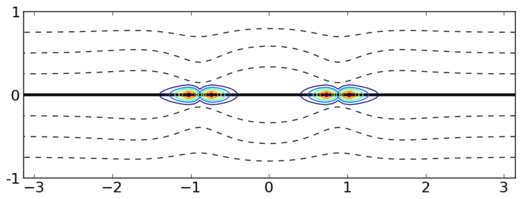} \\
\end{tabular}
\end{center}
\caption{{\bf Critical layers of spanwise localized equilibria of plane Couette flow.}
(a,b) EQ7-1 at $\Rey=1000$ and $30000$ and (c,d) EQ7-2 at $\Rey=1000$ and $30000$
for streamwise wavenumber $\alpha = 1$, with $y$ vertical and $z$ horizontal. Plotting 
conventions are the same as \reffig{f:periodic_pcf_critical}.  The $z \in [-\upi,\upi]$ 
subset of the full $[-3\upi,3\upi]$ computational domain is shown.
}
\label{f:localized_pcf_critical}
\end{figure}

\begin{figure}
\begin{center}
\begin{tabular}{cc}
{\footnotesize (a)} \hspace{-2.5mm} \includegraphics[width=0.45\textwidth]{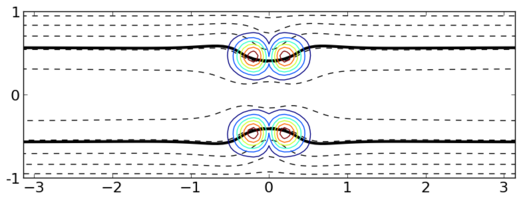} &
{\footnotesize (c)} \hspace{-2.5mm} \includegraphics[width=0.45\textwidth]{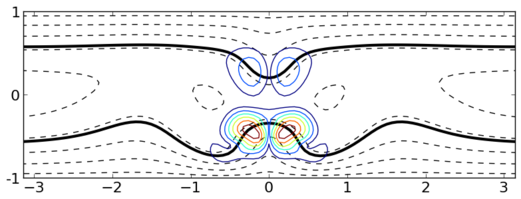} \\
{\footnotesize (b)} \hspace{-2.5mm} \includegraphics[width=0.45\textwidth]{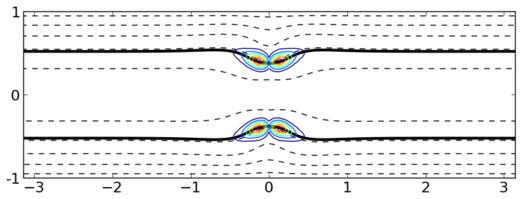} &
{\footnotesize (d)} \hspace{-2.5mm} \includegraphics[width=0.45\textwidth]{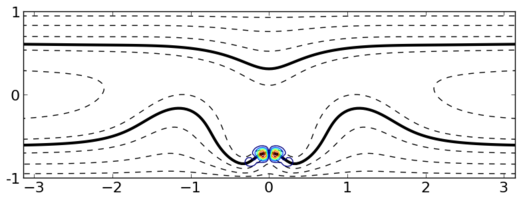} 
\end{tabular}
\end{center}
\caption{{\bf Critical layers of spanwise localized traveling waves of channel flow.}
(a,b) TW1-1 at $\Rey=2000$ and $30000$ and (c,d) TW2-1 at $\Rey=2000$ and $20000$
for streamwise wavenumber $\alpha = 1$, with $y$ vertical and $z$ horizontal. Plotting 
conventions are the same as \reffig{f:periodic_pcf_critical} except total streamwise 
velocity contours are shown at levels $\utot = \{0.1, 0.3, 0.5, 0.7, 0.9\}$. The 
$z \in [-\upi,\upi]$ subset of the full $[-3\upi,3\upi]$ computational domain is shown.
}
\label{f:localized_chflow_critical}
\end{figure}

\refFig{f:localized_pcf_critical} shows that the critical layer structure of EQ7 
carries over directly to its spanwise-localized counterparts EQ7-1 and EQ7-2,
with tapering to laminar flow at large $|z|$. In particular the isolated 
concentrations $\hbu_1$ on the critical layer can be seen, by comparison with 
\reffig{f:couette_crosssections}, to result from the first-harmonic $x$ 
variations of the $y,z$-localized and concentrated vortex structures. 
\refFig{f:localized_chflow_critical}(a,b) shows that EQ7 and EQ7-1's critical 
layer structure carries over to TW1-1, in two copies mirrored symmetrically 
about $y=0$. The channel traveling waves have a nonzero wavespeed $c$ and lack 
EQ7's $\sxy$ symmetry, and therefore have a critical layer $\hutotO(y,z) - c = 0$ 
whose height varies in $z$.

\refFig{f:localized_chflow_critical}(c,d) shows critical layer development 
for the $y$-asymmetric channel traveling wave TW2-1. Note that $y$-asymmetry 
increases while $y$ and $z$ length scales decrease with increasing $\Rey$. In 
particular, $\hbu_1$ concentrates in a smaller region that approaches the wall
as $\Rey$ increases. An obvious question is whether this represents a near-wall 
coherent structure that is constant in wall units. We intend to address this 
question in future work. For the time being we note that the behavior illustrated 
in \reffig{f:localized_chflow_critical}(c,d) is still subject to a prescribed
length scale in the form of the streamwise wavelength $\alpha$, and that 
this prescription must be removed, by streamwise localization or proper 
scaling with $\Rey$, in order for the lengthscales to be determined naturally.

\section{Conclusions}
\label{s:conclusion}

We have found a number of new spanwise-localized equilibrium solutions of plane 
Couette flow and traveling-wave solutions of channel flow, and additionally a few 
spanwise periodic solutions of channel flow incidental to the construction of the 
localized solutions. The spanwise localized solutions consist of a core region that
closely resembles a spanwise periodic solution, a transition region, and exponentially
decaying tails. The decay rate of the tails is $e^{-\alpha |z|}$,  and their structure is 
determined by solely by the streamwise wavenumber, the laminar flow profile, and the wavespeed, and is otherwise 
independent of the structure of the core region. The solutions related to \cite{ItanoPRL09} 
and \cite{GibsonJFM09}'s HSV/EQ7 display clear scale separation and asymptotic scaling 
in streamwise Fourier harmonics, suggesting that they are amenable to analysis via a 
reduced-order PDE retaining only a few harmonics. 

Several solutions, namely TW2-1 and TW2-2, capture particularly isolated and elemental 
exact coherent structures in the near-wall of shear flows, which suggestively resemble 
the lambda vortices that form in spatially developing flows. These solutions consist of
long bands of concentrated vortices near the walls, with alternating orientation, and 
roughly aligned with the streamwise axis but tilting slightly in the spanwise and 
wall-normal directions. The concentrated vortices near the walls are flanked by small 
high-speed streaks very near the walls and otherwise surrounded by very large regions 
where the streamwise velocity is reduced relative to the laminar background. These 
solutions capture, as exact time-independent solutions of Navier-Stokes, the process 
by which near-wall vortices exchange momentum between the near-wall and core regions 
of wall-bounded shear flows and thereby increase drag.

\vspace*{0.6cm}

\noindent{\bf Acknowledgments.}
The authors thank Tobias Schneider, Greg Chini, Spencer Sherwin, and Philip Hall for 
illuminating discussions.

\bibliographystyle{jfm}
\bibliography{hairpin}

\end{document}